\documentclass[5p,times]{elsarticle}
\usepackage{hyperref}

\journal{Knowledge Based Systems}

\usepackage{balance}
\usepackage{listings}
\usepackage{color}

\definecolor{dkgreen}{rgb}{0,0.6,0}
\definecolor{gray}{rgb}{0.5,0.5,0.5}
\definecolor{mauve}{rgb}{0.58,0,0.82}

\makeatletter
\def\lst@makecaption{%
  \def\@captype{table}%
  \@makecaption
}
\makeatother

\newcommand{\vct}[1]{\boldsymbol{#1}} 

\newtheorem{definition}{Definition}

\graphicspath{{img/}}

\usepackage{url}
\usepackage{multirow}

\usepackage{amssymb}
\usepackage{amsmath}

\usepackage{algorithmic}
\usepackage{array}

\usepackage[caption=false,font=footnotesize,labelfont=sf,textfont=sf]{subfig}

\begin{document}

\begin{frontmatter}

\title{Graph Embedding Techniques, Applications, and Performance: A Survey}

\author{Palash~Goyal
		and Emilio~Ferrara}
\address{University of Southern California, Information Sciences Institute\\4676 Admiralty Way, Suite 1001. Marina del Rey, CA. 90292, USA}

\begin{abstract}
Graphs, such as social networks, word co-occurrence networks, and communication networks, occur naturally in various real-world applications. Analyzing them yields insight into the structure of society, language, and different patterns of communication. Many approaches have been proposed to perform the analysis. Recently, methods which use the representation of graph nodes in vector space have gained traction from the research community. In this survey, we provide a comprehensive and structured analysis of various graph embedding techniques proposed in the literature. We first introduce the embedding task and its challenges such as scalability, choice of dimensionality, and features to be preserved, and their possible solutions. We then present three categories of approaches based on factorization methods, random walks, and  deep learning, with examples of representative algorithms in each category and analysis of their performance on various tasks. We evaluate these state-of-the-art methods on a few common datasets and compare their performance against one another. Our analysis concludes by suggesting some potential applications and future directions. We finally present the open-source Python library we developed, named GEM (\textit{Graph Embedding Methods}, available at \url{https://github.com/palash1992/GEM}), which provides all presented algorithms within a unified interface to foster and facilitate research on the topic.
\end{abstract}


\end{frontmatter}


\section{Introduction}

Graph analysis has been attracting increasing attention in the recent years due the ubiquity of networks in the real world. Graphs (a.k.a. networks) have been used to denote information in various areas including biology (Protein-Protein interaction networks)\cite{theocharidis2009network}, social sciences (friendship networks)\cite{freeman2000visualizing} and linguistics (word co-occurrence networks)\cite{i2001small}. Modeling the interactions between entities as graphs has enabled researchers to understand the various network systems in a systematic manner\cite{leskovec2007graph}. For example, social networks have been used for applications like friendship or content recommendation, as well as for advertisement \cite{liben2007link}. Graph analytic tasks can be broadly abstracted into the following four categories: \textit{(a)} node classification\cite{bhagat2011node}, \textit{(b)} link prediction\cite{liben2007link}, \textit{(c)} clustering\cite{ding2001min}, and \textit{(d)} visualization\cite{maaten2008visualizing}. Node classification aims at determining the label of nodes (a.k.a. vertices) based on other labeled nodes and the topology of the network. Link prediction refers to the task of predicting missing links or links that are likely to occur in the future. Clustering is used to find subsets of similar nodes and group them together; finally, visualization helps in providing insights into the structure of the network.


In the past few decades, many methods have been proposed for the tasks defined above. For node classification, there are broadly two categories of approaches --- methods which use random walks to propagate the labels\cite{azran2007rendezvous,baluja2008video}, and methods which extract features from nodes and apply classifiers on them\cite{bhagat2007applying,lu2003link}. Approaches for link prediction include similarity based methods\cite{jaccard1901etude,adamic2003friends}, maximum likelihood models\cite{clauset2008hierarchical,white1976social}, and probabilistic models\cite{friedman1999learning,heckerman2007probabilistic}. Clustering methods include attribute based models\cite{zhou2009graph} and methods which directly maximize (resp., minimize) the inter-cluster (resp., intra-cluster) distances\cite{ding2001min,shi2000normalized}. This survey will provide a taxonomy that captures these application domains and the existing strategies.

Typically, a model defined to solve graph-based problems either operates on the original graph adjacency matrix or on a derived vector space. Recently, the methods based on representing networks in vector space, while preserving their properties, have become widely popular\cite{Ahmed2013,Tang2015,Wang2016}. Obtaining such an embedding is useful in the tasks defined above.\footnote{The term \textit{graph embedding} has been used in the literature in two ways: to represent an entire graph in vector space, or to represent each individual node in vector space. In this paper, we use the latter definition since such representations can be used for tasks like node classification, differently from the former representation.} The embeddings are input as features to a model and the parameters are learned based on the training data. This obviates the need for complex classification models which are applied directly on the graph. 

\subsection{Challenges}
Obtaining a vector representation of each node of a graph is inherently difficult and poses several challenges which have been driving research in this field: 

\textbf{(i) Choice of property}: A ``good'' vector representation of nodes should preserve the structure of the graph and the connection between individual nodes. The first challenge is choosing the property of the graph which the embedding should preserve. Given the plethora of distance metrics and properties defined for graphs, this choice can be difficult and the performance may depend on the application.

\textbf{(ii) Scalability}: Most real networks are large and contain millions of nodes and edges --- embedding methods should be scalable and able to process large graphs. Defining a scalable model can be challenging especially when the model is aimed to preserve global properties of the network.

\textbf{(iii) Dimensionality of the embedding}: Finding the optimal dimensions of the representation can be hard. For example, higher number of dimensions may increase the reconstruction precision but will have high time and space complexity. The choice can also be application-specific depending on the approach: E.g., lower number of dimensions may result in better link prediction accuracy if the chosen model only captures local connections between nodes.

\subsection{Our contribution}
This survey provides a three-pronged contribution:

\textit{(1)} We propose a taxonomy of approaches to graph embedding, and explain their differences. We define four different tasks, \textit{i.e.,} application domains of graph embedding techniques. We illustrate the evolution of the topic, the challenges it faces, and future possible research directions. 

\textit{(2)} We provide a detailed and systematic analysis of various graph embedding models and discuss their performance on the various tasks. For each method, we analyze the properties preserved and its accuracy, through comprehensive comparative evaluation on a few common data sets and application scenarios. Furthermore, we perform hyperparameter sensitivity analysis on the methods evaluated to test their robustness and provide an understanding of the dependence of optimal hyperparameter on the task performed.

\textit{(3)} To foster further research in this topic, we finally present \textit{GEM}, the open-source Python library we developed that provides, under a unified interface,  implementations of \textit{all} graph embedding methods discussed in this survey. To the best of our knowledge, this is the first paper to survey graph embedding techniques and their applications.

\subsection{Organization of the survey}
The survey is organized as follows. In Section \ref{sec:problem}, we provide the definitions required to understand the problem and models discussed next. Section \ref{sec:approaches} proposes a taxonomy of graph embedding approaches and provides a description of representative algorithms in each category. The list of applications for which researchers have used the representation learning approach for graphs is provided in Section \ref{sec:applications}. We then describe our experimental setup (Section \ref{sec:exp_setup}) and evaluate the different models (Section \ref{sec:exp_analysis}). Section \ref{sec:gem} introduces our Python library for graph embedding methods. Finally, in Section \ref{sec:conclusion} we draw our conclusions and discuss potential applications and future research direction.

\section{Definitions and Preliminaries}\label{sec:problem}
We represent the set $\{1, \cdots, n\}$ by $[n]$ in the rest of the paper.
We start by formally defining several preliminaries which have been defined similar to Wang \emph{et al.}~\cite{Wang2016}.

\begin{definition}{(Graph)} A graph $G(V,E)$ is a collection of $V=\{v_1, \cdots, v_n\}$ vertices (a.k.a. nodes) and $E=\{e_{ij}\}_{i,j=1}^n$ edges. The adjacency matrix $S$ of graph $G$ contains non-negative weights associated with each edge: $s_{ij} \geq 0$. If $v_i$ and $v_j$ are not connected to each other, then $s_{ij}=0$. For undirected weighted graphs, $s_{ij}=s_{ji} \hspace{5pt} \forall i,j \in [n]$.
\end{definition}

The edge weight $s_{ij}$ is generally treated as a measure of similarity between the nodes $v_i$ and $v_j$. The higher the edge weight, the more similar the two nodes are expected to be.
\begin{definition}{(First-order proximity)} Edge weights $s_{ij}$ are also called first-order proximities between nodes $v_i$ and $v_j$, since they are the first and foremost measures of similarity between two nodes.
\end{definition}

We can similarly define higher-order proximities between nodes. For instance,
\begin{definition}{(Second-order proximity)} The second-order proximity between a pair of nodes describes the proximity of the pair's neighborhood structure. Let $\vct{s_i} = [s_{i1}, \cdots, s_{in}]$ denote the first-order proximity between $v_i$ and other nodes. Then, second-order proximity between $v_i$ and $v_j$ is determined by the similarity of $\vct{s_i}$ and $\vct{s_j}$.
\end{definition}

Second-order proximity compares the neighborhood of two nodes and treats them as similar if they have a similar neighborhood.
It is possible to define higher-order proximities using other metrics, e.g. Katz Index, Rooted PageRank, Common Neighbors, Adamic Adar, etc. (for detailed definitions, omitted here in the interest of space, see Ou \emph{et al.}~\cite{Ou2016}).
%
%
Next, we define a graph embedding:
\begin{definition}{(Graph embedding)} Given a graph $G=(V,E)$, a graph embedding is a mapping $f:v_i \rightarrow \vct{y_i} \in \mathbb{R}^d \hspace{5pt} \forall i \in [n]$ such that $d \ll |V|$ and the function $f$ preserves some proximity measure defined on  graph $G$. 
\end{definition}

An embedding therefore maps each node to a low-dimensional feature vector and tries to preserve the connection strengths between vertices. 
For instance, an embedding preserving first-order proximity might be obtained by minimizing $\sum_{i,j} s_{ij} \|\vct{y_i} - \vct{y_j}\|_2^2$. 
Let two node pairs $(v_i, v_j)$ and $(v_i, v_k)$ be associated with connections strengths such that $s_{ij} > s_{ik}$. In this case, $v_i$ and $v_j$ will be mapped to points in the embedding space that will be closer each other than the mapping of $v_i$ and $v_k$. 


\section{Algorithmic Approaches: A Taxonomy}\label{sec:approaches}
In the past decade, there has been a lot of research in the field of graph embedding, with a focus on designing new embedding algorithms. More recently, researchers pushed forward scalable embedding algorithms that can be applied on graphs with millions of nodes and edges. In the following, we provide historical context about the research progress in this domain (\S\ref{sub:history}), then propose a taxonomy of graph embedding techniques (\S\ref{sub:taxonomy}) covering \textit{(i)} factorization methods (\S\ref{tax:factorization}), \textit{(ii)} random walk techniques (\S\ref{tax:random_walks}), \textit{(iii)} deep learning (\S\ref{tax:deep}), and \textit{(iv)} other miscellaneous strategies (\S\ref{tax:others}).

\subsection{Graph Embedding Research Context and Evolution}\label{sub:history}
In the early 2000s, researchers developed graph embedding algorithms as part of dimensionality reduction techniques. They would construct a similarity graph for a set of $n$ $D$-dimensional points based on neighborhood and then embed the nodes of the graph in a $d$-dimensional vector space, where $d$ $\ll$ $D$. The idea for embedding was to keep connected nodes closer to each other in the vector space. Laplacian Eigenmaps \cite{belkin2001laplacian} and Locally Linear Embedding (LLE) \cite{Roweis2000} are examples of algorithms based on this rationale. However, scalability is a major issue in this approach, whose time complexity is $O(|V|^2)$.

Since 2010, research on graph embedding has shifted to obtaining scalable graph embedding techniques which leverage the sparsity of real-world networks. For example, Graph Factorization \cite{Ahmed2013} uses an approximate factorization of the adjacency matrix as the embedding. LINE \cite{Tang2015} extends this approach and attempts to preserve both first order and second proximities. HOPE \cite{Ou2016} extends LINE to attempt preserve high-order proximity by decomposing the similarity matrix rather than adjacency matrix using a generalized Singular Value Decomposition (SVD). SDNE \cite{Wang2016} uses autoencoders to embed graph nodes and capture highly non-linear dependencies. The new scalable approaches have a time complexity of $O(|E|)$.

\subsection{A Taxonomy of Graph Embedding Methods}\label{sub:taxonomy}
We propose a taxonomy of embedding approaches. We categorize the embedding methods into three broad categories: (1) Factorization based, (2) Random Walk based, and (3) Deep Learning based. Below we explain the characteristics of each of these categories and provide a summary of a few representative approaches for each category (cf. Table \ref{tab:approaches}), using the notation presented in Table \ref{tab:not}. 

\begin{table*}[!htbp]
	\centering
    \renewcommand{\arraystretch}{1.3}
	\begin{tabular}{| c | c | c | c | c | c |}
         \hline
             \textbf{Category} & \textbf{Year} & \textbf{Published} & \textbf{Method} & \textbf{Time Complexity} & \textbf{Properties preserved}\\ \hline
      \multirow{5}{*}{Factorization} 
      		& 2000 & Science\cite{Roweis2000} & LLE & O($|E|d^2$)&\\ \cline{2-5}
            & 2001 & NIPS\cite{belkin2001laplacian} & Laplacian Eigenmaps & O($|E|d^2$) &$1^{st}$ order proximity\\ \cline{2-5}
            & 2013 & WWW\cite{Ahmed2013} & Graph Factorization & O($|E|d$) &\\ \cline{2-6}
            & 2015 & CIKM\cite{Cao} & GraRep & O($|V|^3$)&\\ \cline{2-5}
            & 2016 & KDD\cite{Ou2016} & HOPE & O($|E|d^2$) & $1-k^{th}$ order proximities\\ \cline{1-5}
      \multirow{3}{*}{Random Walk}
            & 2014 & KDD\cite{Perozzi2014}  & DeepWalk & O($|V|d$) & \\ \cline{2-6}             
            & 2016 & KDD\cite{Grover2016} & \textit{node2vec} & O($|V|d$) & $1-k^{th}$ order proximities,\\ 
            & & & & & structural equivalence\\ \hline
      \multirow{3}{*}{Deep Learning}
            & 2016 & KDD\cite{Wang2016} & SDNE & O($|V||E|$) & $1^{st}$ and $2^{nd}$ order proximities\\ \cline{2-6}
            & 2016 & AAAI\cite{cao2016deep} & DNGR & O($|V|^2$) & $1-k^{th}$ order proximities\\ \cline{2-6}
            & 2017 & ICLR\cite{kipf2016semi} & GCN & O($|E|d^2$) & $1-k^{th}$ order proximities\\ \hline
         Miscellaneous & 2015 & WWW\cite{Tang2015} & LINE & O($|E|d$) & $1^{st}$ and $2^{nd}$ order proximities\\ \hline
	\end{tabular}
    \caption{List of graph embedding approaches}
    \label{tab:approaches}
\end{table*}

\begin{table}[!htbp]
	\footnotesize
	\centering
    \renewcommand{\arraystretch}{1.3}
        \begin{tabular}{|@{}c@{}|c|}
            \hline
             $G$ & Graphical representation of the data \\ \hline
             $V$ & Set of vertices in the graph \\ \hline
             $E$ & Set of edges in the graph \\ \hline
             $d$ & Number of dimensions \\ \hline
             $Y$ & Embedding of the graph, $|V| \times d$ \\ \hline
             $Y_i$ & Embedding of node $v_i$, $1 \times d$ (also $i^{th}$ row of $Y$) \\ \hline
             $Y_s$ & Source embedding of a directed graph, $|V| \times d$ \\ \hline
             $Y_t$ & Target embedding of a directed graph, $|V| \times d$ \\ \hline
             $W$ & Adjacency matrix of the graph, $|V| \times |V|$ \\ \hline
             $D$ & Diagonal matrix of the degree of each vertex,  $|V| \times |V|$  \\ \hline
             $L$ & Graph Laplacian $(L=D-W)$, $|V| \times |V|$  \\ \hline
             $<Y_i, Y_j>$ & Inner product of $Y_i$ and $Y_j$ i.e. $Y_i Y_j^T$ \\ \hline
             $S$ & Similarity matrix of the graph, $|V| \times |V|$ \\ \hline
        \end{tabular}
    \caption{Summary of notation}
    \label{tab:not}
\end{table}
    
\subsection{Factorization based Methods} \label{tax:factorization}
Factorization based algorithms represent the connections between nodes in the form of a matrix and factorize this matrix to obtain the embedding. The matrices used to represent the connections include node adjacency matrix, Laplacian matrix, node transition probability matrix, and Katz similarity matrix, among others. Approaches to factorize the representative matrix vary   based on the matrix properties. If the obtained matrix is positive semidefinite, e.g. the Laplacian matrix, one can use eigenvalue decomposition. For unstructured matrices, one can use gradient descent methods to obtain the embedding in linear time. 

\subsubsection{Locally Linear Embedding (LLE)}
LLE \cite{Roweis2000} assumes that every node is a linear combination of its neighbors in the embedding space. If we assume that the adjacency matrix element $W_{ij}$ of graph $G$ represents the weight of node $j$ in the representation of node $i$, we define
\begin{equation*}
    Y_i \approx \sum_j W_{ij}Y_j \hspace{16pt} \forall i \in V.
\end{equation*}
Hence, we can obtain the embedding $Y^{N\times d}$ by minimizing 
\begin{equation*}
    \phi(Y) = \sum_i |Y_i - \sum_j W_{ij}Y_j|^2,
\end{equation*}
To remove degenerate solutions, the variance of the embedding is constrained as $\frac{1}{N} Y^T Y = I$. To further remove translational invariance, the embedding is centered around zero: $\sum_i Y_i = 0$.
The above constrained optimization problem can be reduced to an eigenvalue problem, whose solution is to take the bottom $d +1$ eigenvectors of the sparse matrix $(I-W)^T (I-W)$  and discarding the eigenvector corresponding to the smallest eigenvalue.

\subsubsection{Laplacian Eigenmaps}
Laplacian Eigenmaps \cite{belkin2001laplacian} aims to keep the embedding of two nodes close when the weight $W_{ij}$ is high. Specifically, they minimize the following objective function
\begin{equation*}
    \begin{split}
        \phi(Y) &= \frac{1}{2}\sum_{i,j}|Y_i - Y_j|^2 W_{ij}\\
                &= tr(Y^T L Y),\\
    \end{split}
\end{equation*}
where $L$ is the Laplacian of  graph $G$. The objective function is subjected to the constraint $Y^T D Y = I$ to eliminate trivial solution. The solution to this can be obtained by taking the eigenvectors corresponding to the $d$ smallest eigenvalues of the normalized Laplacian, $L_{norm} = D^{-1/2}LD^{-1/2}$.

\subsubsection{Cauchy Graph Embedding}
Laplacian Eigenmaps uses a quadratic penalty function on the distance between embeddings. The objective function thus emphasizes preservation of dissimilarity between nodes more than their similarity. This may yield embeddings which do not preserve local topology, which can be defined as the equality between relative order of edge weights ($W_{ij}$) and inverse order of distances in the embedded space ($|Y_i - Y_j|^2$). Cauchy Graph Embedding \cite{luo2011cauchy} tackles this problem by replacing the quadratic function $|Y_i - Y_j|^2$ with $\frac{|Y_i - Y_j|^2}{|Y_i - Y_j|^2 + \sigma^2}$. Upon rearrangement, the objective function to be maximized becomes
\begin{equation*}
	\phi(Y) = \sum_{i,j}\frac{W_{ij}}{|Y_i - Y_j|^2 + \sigma^2}, \\
\end{equation*}
with constraints $Y^T Y = I$ and $\sum_i Y_i = 0$ for each $i$. The new objective is an inverse function of distance and thus puts emphasis on similar nodes rather than dissimilar nodes. The authors propose several variants including Gaussian, Exponential and Linear embeddings with varying relative emphasis on the distance between nodes.

\subsubsection{Structure Preserving Embedding (SPE)}
Structure Preserving Embedding (\cite{shaw2009structure}) is another approach which extends Laplacian Eigenmaps. SPE aims to reconstruct the input graph exactly. The embedding is stored as a positive semidefinite kernel matrix $K$ and a connectivity algorithm $\mathcal{G}$ is defined which reconstructs the graph from $K$. The kernel $K$ is chosen such that it maximizes $tr(KW)$ which attempts to recover rank-1 spectral embedding. Choice of the connectivity algorithm $\mathcal{G}$ induces constraints on this objective function. For e.g., if the connectivity scheme is to connect each node to neighbors which lie within a ball of radius $\epsilon$,  the constraint $(K_{ii} + K_{jj} - 2K_{ij})(W_{ij} - 1/2) \leq \epsilon (W_{ij} - 1/2)$ produces a kernel which can perfectly reconstruct the original graph. To handle noise in the graph, a slack variable is added. For $\xi$-connectivity, the optimization thus becomes $max\ tr(KA) - C\xi$ s.t. $(K_{ii} + K_{jj} - 2K_{ij})(W_{ij} - 1/2) \leq \epsilon (W_{ij} - 1/2) - \xi$, where $\xi$ is the slack variable and $C$ controls slackness.

\subsubsection{Graph Factorization (GF)}
To the best of our knowledge, Graph Factorization \cite{Ahmed2013} was the first method to obtain a graph embedding in $O(|E|)$ time. To obtain the embedding, GF factorizes the adjacency matrix of the graph,  minimizing the following loss function
\begin{equation*}
    \phi(Y, \lambda) = \frac{1}{2}\sum_{(i,j)\in E}(W_{ij} - <Y_i, Y_j>)^2 + \frac{\lambda}{2}\sum_i \|Y_i\|^2,
\end{equation*}
where $\lambda$ is a regularization coefficient. Note that the summation is over the observed edges as opposed to all possible edges. This is an approximation in the interest of scalability, and as such it may introduce noise in the solution. Note that as the adjacency matrix is often not positive semidefinite, the minimum of the loss function is greater than 0 even if the dimensionality of embedding is $|V|$.

\subsubsection{GraRep}
GraRep \cite{Cao} defines the node transition probability as $T = D^{-1} W$ and preserves $k$-order proximity by minimizing $\|X^k - Y^k_s Y^{kT}_t\|_F^2$ where $X^k$ is derived from $T^k$ (refer to \cite{Cao} for a detailed derivation). It then concatenates $Y^k_s$ for all $k$ to form $Y_s$. Note that this is similar to HOPE \cite{Ou2016} which minimizes $\|S - Y_s Y_t^T\|_F^2$ where $S$ is an appropriate similarity matrix. The drawback of GraRep is scalability, since $T^k$ can have $O(|V|^2)$ non-zero entries.

\subsubsection{HOPE}
HOPE \cite{Ou2016} preserves higher order proximity by minimizing $\|S - Y_s Y_t^T\|_F^2$, where $S$ is the similarity matrix. The authors experimented with different similarity measures, including Katz Index, Rooted Page Rank, Common Neighbors, and Adamic-Adar score. They represented each similarity measure as $S = M_g^{-1} M_l$, where both $M_g$ and $M_l$ are sparse. This enables HOPE to use generalized Singular Value Decomposition (SVD) \cite{VanLoan1976} to obtain the embedding efficiently.

\subsubsection{Additional Variants}
For the purpose of dimensionality reduction of high dimensional data, there are several other methods developed capable of performing graph embedding. Yan \emph{et al.} \cite{yan2007graph} survey a list of such methods including Principal Component Analysis (PCA)~\cite{jolliffe1986principal}, Linear Discrimant Analysis (LDA)~\cite{martinez2001pca}, ISOMAP~\cite{tenenbaum2000global}, Multidimesional Scaling (MDS)~\cite{kruskal1978multidimensional}, Locality Preserving Properties (LPP)~\cite{he2004locality} and Kernel Eigenmaps~\cite{brand2003continuous}. Matrinex \emph{et al.} \cite{yangnge}) proposed a general framework, non-negative graph embedding, which yields non-negative embeddings for these algorithms.

A number of recent techniques have focused on jointly learning network structure and additional node attribute information available for the network., Augmented Relation Embedding (ARE)~\cite{lin2005semantic} augments network with content based features for images and modifies the graph-Laplacian to capture such information. Text-associated DeepWalk (TADW)~\cite{yang2015network} performs matrix factorization on node similarity matrix disentangling the representation using text feature matrix. Heterogeneous Network Embedding (HNE)~\cite{chang2015heterogeneous} learns representation for each modality of the network and then unifies them into a common space using linear transformations. Other works (\cite{tu2016max, zhang2016homophily, huang2017label}) perform similar transformations between various node attributes and learn joint embedding.

\subsection{Random Walk based Methods} \label{tax:random_walks}
Random walks have been used to approximate many properties in the graph including node centrality\cite{newman2005measure} and similarity\cite{fouss2007random}. They are especially useful when one can either only partially observe the graph, or the graph is too large to measure in its entirety. Embedding techniques using random walks on graphs to obtain node representations have been proposed: \textit{DeepWalk} and \textit{node2vec} are two  examples.

\subsubsection{DeepWalk}
DeepWalk \cite{Perozzi2014} preserves higher-order proximity between nodes by maximizing the probability of observing the last $k$ nodes and the next $k$ nodes in the random walk centered at $v_i$, i.e. maximizing 
    $\log Pr({v_{i-k}, \ldots, v_{i-1}, v_{i+1}, \ldots, v_{i+k}} | Y_i)$,
where $2k + 1$ is the length of the random walk. The model generates multiple random walks each of length $2k + 1$ and performs the optimization over sum of log-likelihoods for each random walk. A dot-product based decoder is used to reconstruct the edges from the node embeddings.

\subsubsection{\textit{node2vec}}
Similar to DeepWalk \cite{Perozzi2014}, \textit{node2vec} \cite{Grover2016}  preserves higher-order proximity between nodes by maximizing the probability of occurrence of subsequent nodes in fixed length random walks.
The crucial difference from DeepWalk is that \textit{node2vec} employs biased-random walks that provide a trade-off between breadth-first (BFS) and depth-first (DFS) graph searches, and hence produces higher-quality and more informative embeddings than DeepWalk. Choosing the right balance enables \textit{node2vec} to preserve community structure as well as structural equivalence between nodes.

\subsubsection{Hierarchical Representation Learning for Networks (HARP)}
DeepWalk and \textit{node2vec} initialize the node embeddings randomly for training the models. As their objective function is non-convex, such initializations can be stuck in local optima. HARP \cite{chen2017harp} introduces a strategy to improve the solution and avoid local optima by better weight initialization. To this purpose, HARP creates hierarchy of nodes by aggregating nodes in the previous layer of hierarchy using graph coarsening. It then generates embedding of the coarsest graph and initializes the node embeddings of the refined graph (one up in the hierarchy) with the learned embedding. It propagates such embeddings through the hierarchy to obtain the embeddings of the original graph. Thus HARP can be used in conjunction with random walk based methods like DeepWalk and \textit{node2vec} to obtain better solutions to the optimization function.

\subsubsection{Walklets}
DeepWalk and \textit{node2vec} implicitly preserve higher order proximity between nodes by generating multiple random walks which connect nodes at various distances due to its stochastic nature. On the other hand, factorization based approches like GF and HOPE explicitly preserve distances between nodes by modeling it in their objective function. Walklets \cite{perozzi2016walklets} combine this idea of explicit modeling with random walks. The model modifies the random walk strategy used in DeepWalk by skipping over some nodes in the graph. This is performed for multiple skip lengths, analogous to factorizing $A^k$ in GraRep, and the resulting set of random walks are used for training the model similar to DeepWalk.

\subsubsection{Additional Variants}
There have been several variations of the above methods proposed recently. Similar to augmenting graph structure with node attributes for factorization based methods, GenVector~\cite{yang2016multi}, Discriminative Deep Random Walk (DDRW)~\cite{li2016discriminative}, Tri-party Deep Network Representation (TriDNR)~\cite{pan2016tri} and \cite{yang2016revisiting} extend random walks to jointly learn network structure and node attributes.

\subsection{Deep Learning based Methods} \label{tax:deep}
The growing research on deep learning 
has led to a deluge of deep neural networks based methods applied to graphs\cite{Wang2016,cao2016deep,niepert2016learning}. Deep autoencoders have been used for dimensionality reduction\cite{bengio2013representation} due to their ability to model non-linear structure in the data. Recently, SDNE \cite{Wang2016}, DNGR \cite{cao2016deep} utilized this ability of deep autoencoder to generate an embedding model that can capture non-linearity in graphs.

\subsubsection{Structural Deep Network Embedding (SDNE)}
Wang \emph{et al.} \cite{Wang2016} proposed to use deep autoencoders to preserve the first and second order network proximities. They achieve this by jointly optimizing the two proximities. 
The approach uses highly non-linear functions to obtain the embedding. The model consists of two parts: unsupervised and supervised. The former consists of an autoencoder aiming at finding an embedding for a node which can reconstruct its neighborhood. The latter is based on Laplacian Eigenmaps\cite{belkin2001laplacian} which apply a penalty when similar vertices are mapped far from each other in the embedding space.

\subsubsection{Deep Neural Networks for Learning Graph Representations (DNGR)}
DNGR combines random surfing with deep autoencoder. The model consists of 3 components: random surfing, \textit{positive pointwise mutual information} (PPMI) calculation and stacked denoising autoencoders. Random surfing model is used on the input graph to generate a probabilistic co-occurence matrix, analogous to similarity matrix in HOPE. The matrix is transformed to a PPMI matrix and input into a stacked denoising autoencoder to obtain the embedding. Inputting PPMI matrix ensures that the autoencoder model can capture higher order proximity. Furthermore, using stacked denoising autoencoders aids robustness of the model in presence of noise in the graph as well as in capturing underlying structure required for tasks such as link prediction and node classification.

\subsubsection{Graph Convolutional Networks (GCN)}
Deep neural network based methods discussed above, namely SDNE and DNGR, take as input the global neighborhood of each node (a row of PPMI for DNGR and adjacency matrix for SDNE). This can be computationally expensive and inoptimal for large sparse graphs. Graph Convolutional Networks (GCNs) \cite{kipf2016semi} tackle this problem by defining a convolution operator on graph. The model iteratively aggregates the embeddings of neighbors for a node and uses a function of the obtained embedding and its embedding at previous iteration to obtain the new embedding. Aggregating embedding of only local neighborhood makes it scalable and multiple iterations allows the learned embedding of a node to characterize global neighborhood.

Several recent papers (\cite{bruna2013spectral, henaff2015deep, duvenaud2015convolutional, li2015gated, defferrard2016convolutional, hamilton2017inductive}) have proposed methods using convolution on graphs to obtain semi-supervised embedding, which can be used to obtain unsupervised embedding by defining unique labels for each node. The approaches vary in the construction of convolutional filters which can broadly be categorized into spatial and spectral filters. Spatial filters operate directly on the original graph and adjacency matrix whereas spectral filters operate on the spectrum of graph-Laplacian.

\subsubsection{Variational Graph Auto-Encoders (VGAE)}
Kipf \emph{et al.} \cite{kipf2016variational} evaluate the performance of variational autoencoders \cite{kingma2013auto} on the task of graph embedding. The model uses a graph convolutional network (GCN) encoder and an inner product decoder. The input is adjacency matrix and they rely on GCN to learn the higher order dependencies between nodes. They empirically show that using variational autoencoders can improve performance compared to non-probabilistic autoencoders.

\subsection{Other Methods} \label{tax:others}
\subsubsection{LINE}
LINE \cite{Tang2015} explicitly defines two functions, one each for first- and second-order proximities, and minimizes the combination of the two. The function for first-order proximity is similar to that of Graph Factorization (GF) \cite{Ahmed2013} in that they both aim to keep the adjacency matrix and dot product of embeddings close. The difference is that GF does this by directly minimizing the difference of the two. Instead, LINE defines two joint probability distributions for each pair of vertices, one using adjancency matrix and the other using the embedding. Then, LINE minimizes the Kullback-Leibler (KL) divergence of these two distributions. The two distributions and the objective function are as follows
\begin{equation*}
    \begin{aligned}
        &p_1(v_i, v_j) = \frac{1}{1 + exp(-<Y_i, Y_j>)}\\
        &\hat{p_1}(v_i, v_j) = \frac{W_{ij}}{\sum_{(i,j) \in E} W_{ij}}\\
        &O_1 = KL(\hat{p_1}, p_1)\\
        &O_1= - \sum_{(i,j) \in E} W_{ij} \log p_1(v_i, v_j).\\
    \end{aligned}
\end{equation*}

The authors similarly define probability distributions and objective function for the second-order proximity.

\subsection{Discussion}
\begin{figure*}[!ht]
	\centering
	\subfloat[Graph $G_1$]{\label{fig_m2} \includegraphics[width=0.32\textwidth,height=0.22\textwidth]{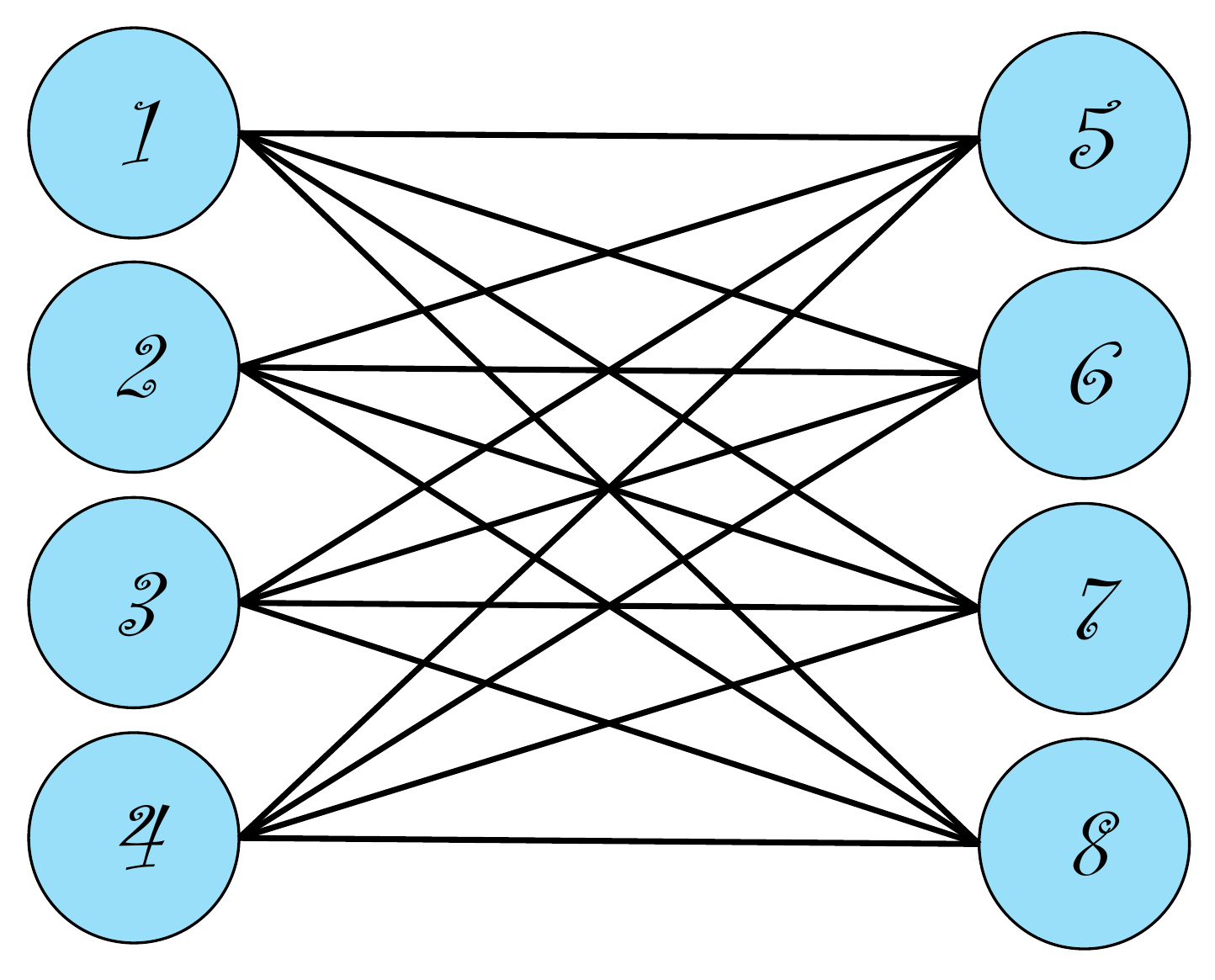}}
	\hfil
	\subfloat[CPE for $G_1$]{\label{fig_m4} \includegraphics[width=0.32\textwidth]{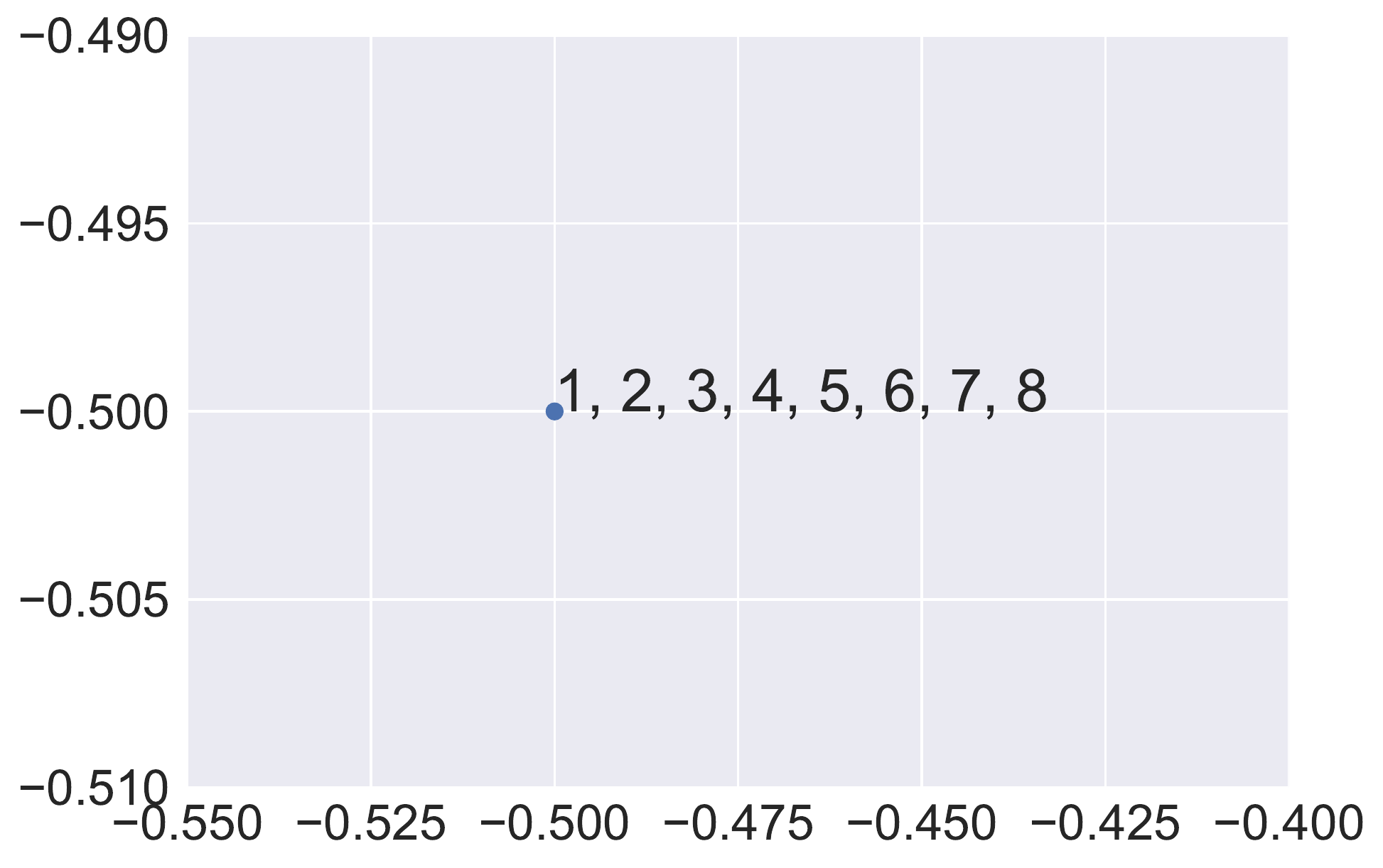}}
	\hfil
	\subfloat[SPE for $G_1$]{\label{fig_m6} \includegraphics[width=0.32\textwidth]{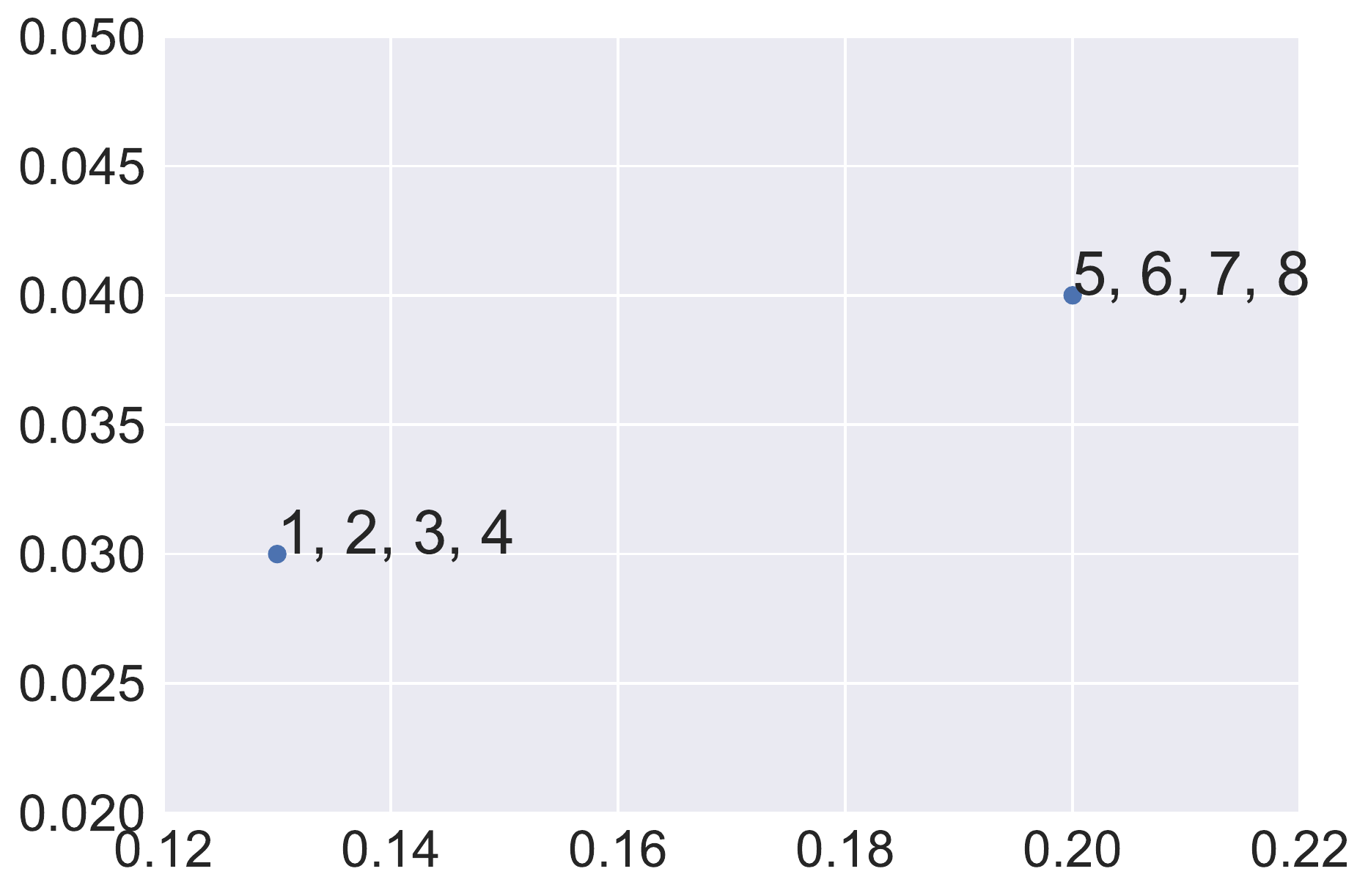}}
    \hfil
    \subfloat[Graph $G_2$]{\label{fig_m2} \includegraphics[width=0.32\textwidth,height=0.22\textwidth]{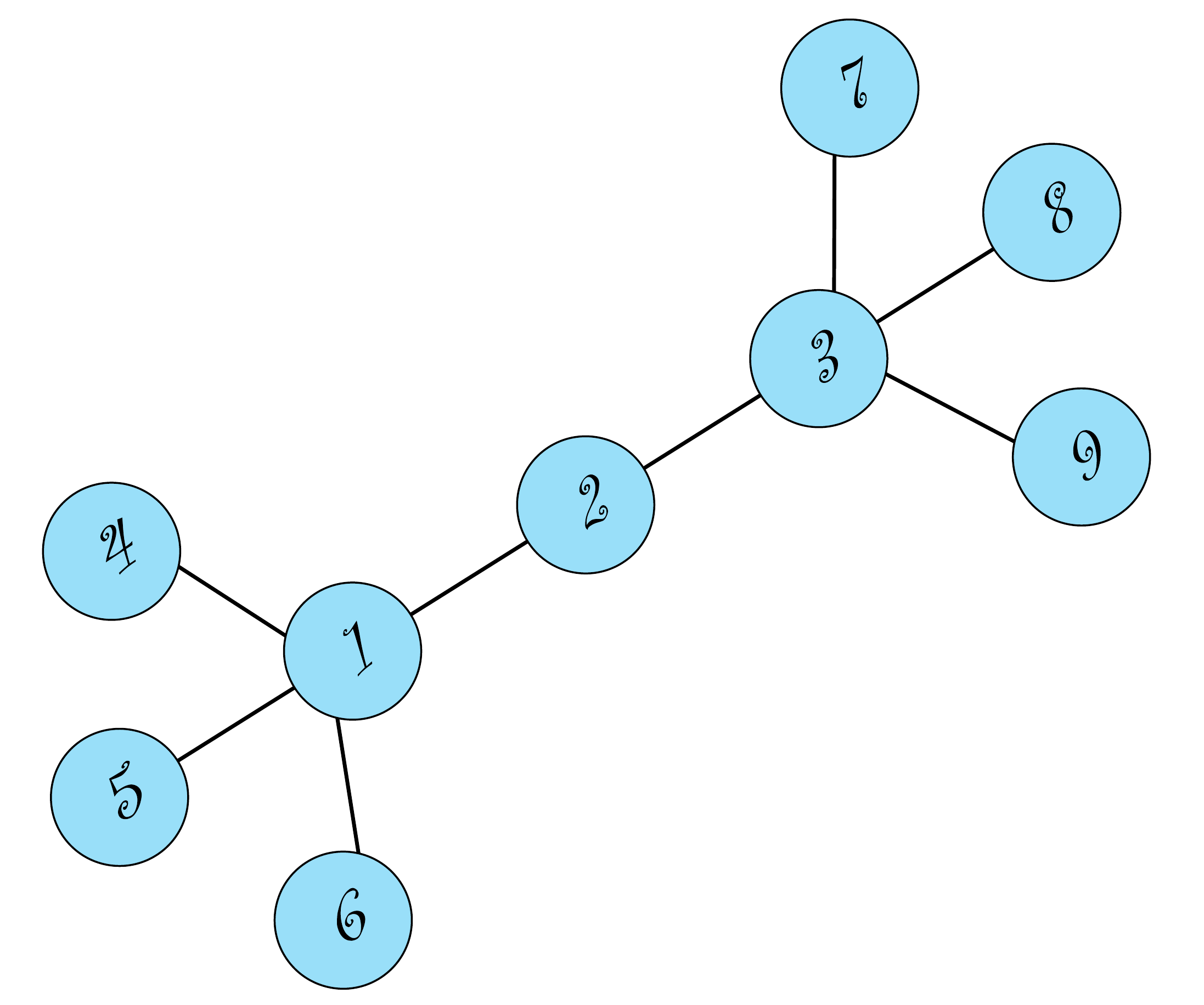}}
	\hfil
    \subfloat[CPE for $G_2$]{\label{fig_m2} \includegraphics[width=0.32\textwidth]{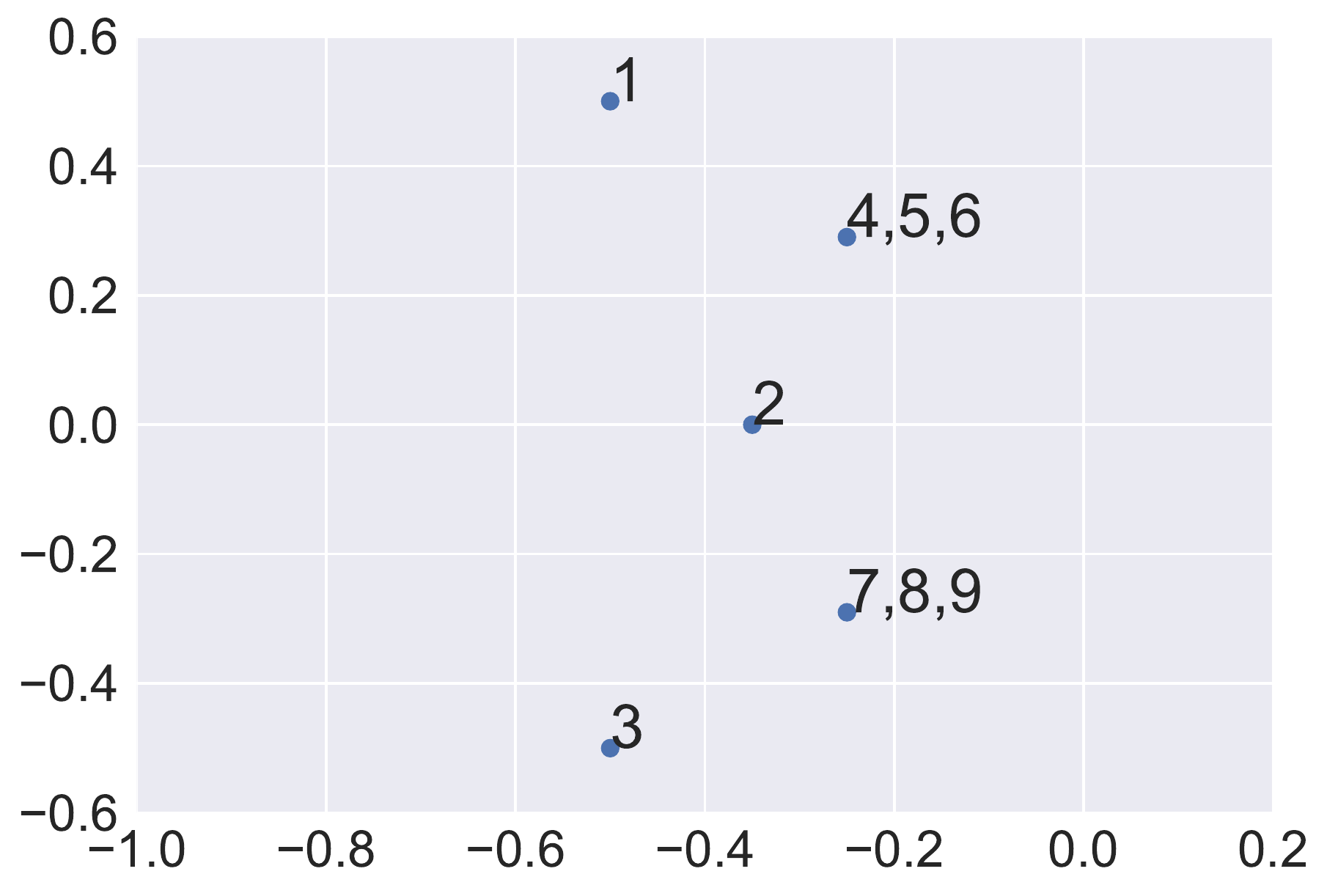}}
	\hfil
    \subfloat[SPE for $G_2$]{\label{fig_m2} \includegraphics[width=0.32\textwidth]{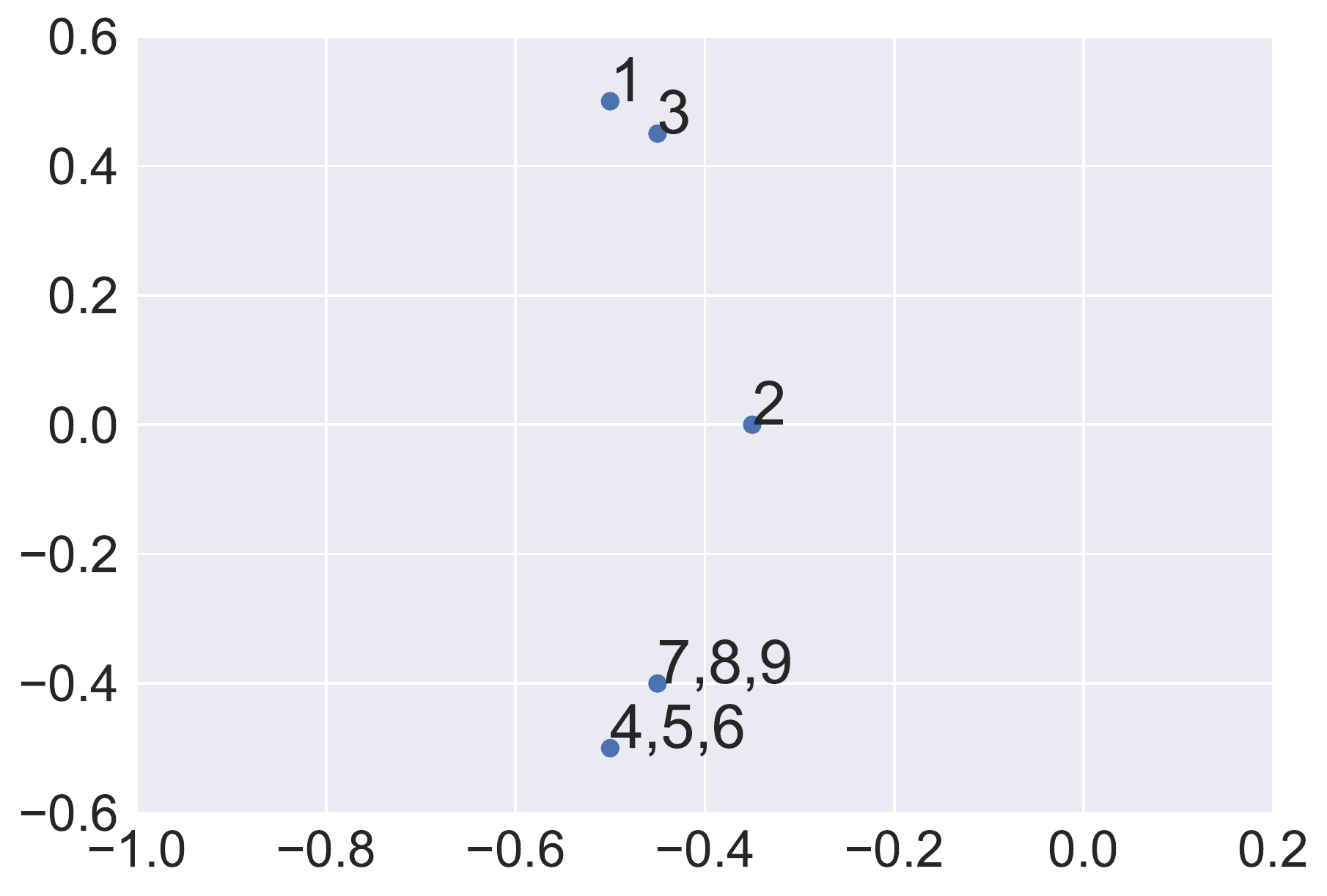}}
	\hfil
	\caption{Examples illustrating the effect of type of similarity preserved. Here, CPE and SPE stand for \textit{Community Preserving Embedding} and \textit{Structural-equivalence Preserving Embedding}, respectively.}
	\label{fig:sim}
\end{figure*}

We can interpret embeddings as representations which describe graph data.
Thus, embeddings can yield insights into the properties of a network.
We illustrate this in Figure \ref{fig:sim}.
Consider a complete bipartite graph $G$.
An embedding algorithm which attempts to keep two connected nodes close (i.e., preserve the community structure), would fail to capture the structure of the graph --- as shown in \ref{fig:sim}(b).
However, an algorithm which embeds structurally-equivalent nodes together learns an interpretable embedding --- as shown in \ref{fig:sim}(c).
Similarly, in \ref{fig:sim}(d) we consider a graph with two star components connected through a hub.
Nodes 1 and 3 are structurally equivalent (they link to the same nodes) and are clustered together in \ref{fig:sim}(f), whereas in \ref{fig:sim}(e) they are far apart.
The classes of algorithms above can be described in terms of their ability to explain the properties of graphs.

Factorization-based methods are not capable of learning an arbitrary function, e.g., to explain network connectivity.
Thus, unless explicitly included in their objective function, they cannot learn structural equivalence.
In random walk based methods, the mixture of equivalences can be controlled to a certain extent by varying the random walk parameters. 
Deep learning methods can model a wide range of functions following the universal approximation theorem \cite{hornik1990universal}: given enough parameters, they can learn the mix of community and structural equivalence, to embed the nodes such that the reconstruction error is minimized.
We can interpret the weights of the autoencoder as a representation of the structure of the graph.
For example, \ref{fig:sim}(c) plots the embedding learned by SDNE for the complete bipartite graph $G_1$.
The autoencoder stored the bipartite structure in weights and achieved perfect reconstruction.
Given the variety of properties of real-world graphs, using general non-linear models that span a large class of functions is a promising direction that warrants further exploration.

\section{Applications}\label{sec:applications}
As graph representations, embeddings can be used in a variety of tasks. 
These applications can be broadly classified as: network compression (\S\ref{sub:netcomp}), visualization (\S\ref{sub:vis}), clustering (\S\ref{sub:clust}), link prediction (\S\ref{sub:linkpred}), and node classification (\S\ref{sub:nodeclass}).

\subsection{Network Compression} \label{sub:netcomp}
Feder \emph{et al.}\cite{feder1991clique} introduced the concept of network compression (a.k.a. graph simplification). 
For a graph $G$, they defined a compression $G^*$ which has smaller number of edges.
The goal was to store the network more efficiently and run graph analysis algorithms faster.
They obtained the compression graph by partitioning the original graph into bipartite cliques and replacing them by trees, thus reducing the number of edges.
Over the years, many researchers have used aggregation based methods \cite{pardalos1994maximum,tian2008efficient,toivonen2011compression} to compress graphs.
The main idea in this line of work is to exploit the link structure of the graph to group nodes and edges.
Navlakha \emph{et al.}\cite{navlakha2008graph} used Minimum Description Length (MDL) \cite{rissanen1978modeling} from information theory to summarize a graph into a graph summary and edge correction.

Similar to these representations, graph embedding can also be interpreted as a summarization of graph.
Wang \emph{et al.}\cite{Wang2016} and Ou \emph{et al.}\cite{Ou2016} tested this hypothesis explicitly by reconstructing the original graph from the embedding and evaluating the reconstruction error.
They show that a low dimensional representation for each node (in the order of 100s) suffices to reconstruct the graph with high precision.

\subsection{Visualization} \label{sub:vis}
Application of visualizing graphs can be dated back to 1736 when Euler used it to solve "Konigsberger Bruckenproblem" \cite{jungnickel2005graphs}.
In the recent years, graph visualization has found applications in software engineering \cite{gansner2000open}, electrical circuits \cite{di1994algorithms}, biology \cite{theocharidis2009network} and sociology \cite{freeman2000visualizing}.
Battista \emph{et al.}\cite{di1994algorithms} and Eades \emph{et al.}\cite{eades1989draw} survey a range of methods used to draw graphs and define aesthetic criteria for this purpose.
Herman \emph{et al.}\cite{herman2000graph} generalize this and view it from an information visualization perspective.
They study and compare various traditional layouts used to draw graphs including tree-, 3D- and hyperbolic-based layouts.

As embedding represents a graph in a vector space, dimensionality reduction techniques like Principal Component Analysis (PCA) \cite{pearson1901liii} and \textit{t-distributed stochastic neighbor embedding} (t-SNE) \cite{maaten2008visualizing} can be applied on it to visualize the graph.
The authors of DeepWalk \cite{Perozzi2014} illustrated the goodness of their embedding approach by visualizing the Zachary's Karate Club network.
The authors of LINE \cite{Tang2015} visualized the DBLP co-authorship network, and showed that LINE is able to cluster  together authors in the same field.
The authors of SDNE \cite{Wang2016} applied it on 20-Newsgroup document similarity network to obtain clusters of documents based on topics.

\subsection{Clustering} \label{sub:clust}
Graph clustering (a.k.a., network partitioning) can be of two types: (a) structure based, and (b) attribute based clustering. 
The former can be further divided into two categories, namely community based, and structurally equivalent clustering. 
Structure-based methods \cite{ding2001min,shi2000normalized,newman2004finding}, aim to find dense subgraphs with high number of intra-cluster edges, and low number of inter-cluster edges.
Structural equivalence clustering \cite{xu2007scan}, on the contrary, is designed to identify nodes with similar roles (like bridges and outliers). 
Attribute based methods \cite{zhou2009graph} utilize node labels, in addition to observed links, to cluster nodes.

White \emph{et al.}\cite{white2005spectral} used $k$-means on the embedding to cluster the nodes and visualize the clusters obtained on Wordnet and NCAA data sets verifying that the clusters obtained have intuitive interpretation.
Recent methods on embedding haven't explicitly evaluated their models on this task and thus it is a promising field of research in the graph embedding community.

\subsection{Link Prediction} \label{sub:linkpred}
Networks are constructed from the observed interactions between entities, which may be incomplete or inaccurate. 
The challenge often lies in identifying spurious interactions and predicting missing information. 
Link prediction refers to the task of predicting  either missing interactions or links that may appear in the future in an evolving network.
Link prediction is pervasive in biological  network analysis, where verifying the existence of links between nodes requires costly experimental tests.
Limiting the experiments to links ordered by presence likelihood has been shown to be very cost effective.
In social networks, link prediction is used to predict probable friendships, which can be used for recommendation and lead to a more satisfactory user experience.
Liben-Nowell \emph{et al.}\cite{liben2007link}, Lu \emph{et al.}\cite{lu2011link} and Hasan \emph{et al.}\cite{al2011survey} survey the recent progress in this field and categorize the algorithms into (a) similarity based (local and global) \cite{jaccard1901etude,adamic2003friends,katz1953new}, (b) maximum likelihood based \cite{clauset2008hierarchical,white1976social} and (c) probabilistic methods \cite{friedman1999learning,heckerman2007probabilistic,yu2006stochastic}.

Embeddings capture inherent dynamics of the network either explicitly or implicitly thus enabling application to link prediction.
Wang \emph{et al.}\cite{Wang2016} and Ou \emph{et al.}\cite{Ou2016} predict links from the learned node representations on publicly available collaboration and social networks. 
In addition, Grover \emph{et al.}\cite{Grover2016} apply it to biology networks.
They show that on these data sets links predicted using embeddings are more accurate than traditional similarity based link prediction methods described above.

\subsection{Node Classification} \label{sub:nodeclass}
Often in networks, a fraction of nodes are labeled. 
In social networks, labels may indicate interests, beliefs, or demographics.
In language networks, a document may be labeled with topics or keywords, whereas the labels of entities in biology networks may be based on  functionality.
Due to various factors, labels may be unknown for large fractions of nodes.
For example, in social networks many users do not provide their demographic information due to privacy concerns.
Missing labels can be inferred using the labeled nodes and the links in the network.
The task of predicting these missing labels is also known as node classification.
Bhagat \emph{et al.}\cite{bhagat2011node} survey the methods used in the literature for this task.
They classify the approaches into two categories, i.e., feature extraction based and random walk based.
Feature-based models \cite{bhagat2007applying,lu2003link,neville2000iterative} generate features for nodes based on their neighborhood and local network statistics and then apply a classifier like Logistic Regression \cite{hosmer2013applied} and  Naive Bayes \cite{mccallum1998comparison} to predict the labels.
Random walk based models \cite{azran2007rendezvous,baluja2008video} propagate the labels with random walks.

Embeddings can be interpreted as automatically extracted node features based on network structure and thus falls into the first category.
Recent work\cite{Perozzi2014,Tang2015,Ou2016,Wang2016,Grover2016} has evaluated the predictive power of embedding on various information networks including language, social, biology and collaboration graphs.
They show that embeddings can predict missing labels with high precision.

\section{Experimental Setup}\label{sec:exp_setup}
Our experiments evaluate the feature representations obtained using the methods reviewed before on the previous four application domains.
Next, we specify the datasets and evaluation metrics we used. 
The experiments were performed on a Ubuntu 14.04.4 LTS system with 32 cores, 128 GB RAM and a clock speed of 2.6 GHz. The GPU used for deep network based models was Nvidia Tesla K40C.
\subsection{Datasets}
\begin{table*}[!htbp]
	\centering\footnotesize
    \renewcommand{\arraystretch}{1.3}
  \begin{tabular}{| c | c | c | c | c | c | c | c |}
           \hline      
            & \textbf{Synthetic} & \multicolumn{3}{c|}{\textbf{Social Network}} & \multicolumn{2}{c|}{\textbf{Collaboration Network}} & \textbf{Biology Network}\\ \hline
      Name & SYN-SBM & KARATE & BLOGCATALOG & YOUTUBE & HEP-TH & ASTRO-PH & PPI\\ \hline
      $|V|$ & 1024 & 34 & 10,312 & 1,157,827 & 7,980 & 18,772 & 3,890\\ \hline
      $|E|$ & 29,833 & 78 & 333,983 & 4,945,382 & 21,036 & 396,160 & 38,739\\ \hline
      Avg. degree & 58.27 & 4.59 & 64.78 & 8.54 & 5.27 & 31.55 & 19.91\\ \hline
      No. of labels & 3 & 4 & 39 & 47 & - & - & 50\\ \hline
  \end{tabular}
  \caption{Dataset Statistics}
  \label{tab:data_summ}
\end{table*}

We evaluate the embedding approaches on a synthetic and 6 real datasets. 
The datasets are summarized in Table ~\ref{tab:data_summ}.

\textbf{SYN-SBM}: We generate synthetic graph using Stochastic Block Model \cite{Yuchung1987} with 1024 nodes and 3 communities. 
We set the in-block and cross-block probabilities as 0.1 and 0.01 respectively. 
As we know the community structure in this graph, we use it to visualize the embeddings learnt by various approaches.

\textbf{KARATE} \cite{zachary1977information}: Zachary's karate network is a well-known social network of a university karate club.  
It has been widely studied in social network analysis.
The network has 34 nodes, 78 edges and  2 communities.

\textbf{BLOGCATALOG} \cite{tang2009relational}: This is a network of social relationships of the bloggers listed on the BlogCatalog website. 
The labels represent blogger interests inferred through the metadata provided by the bloggers. 
The network has 10,312 nodes, 333,983 edges and 39 different labels.

\textbf{YOUTUBE} \cite{tang2009scalable}: This is a social network of Youtube users.
This is a large network containing 1,157,827 nodes and 4,945,382 edges.
The labels represent groups of users who enjoy common video genres.

\textbf{HEP-TH} \cite{Gehrke2003}: The original dataset contains abstracts of papers in \textit{High Energy Physics Theory} for the period from January 1993 to April 2003.
We create a collaboration network for the papers published in this period.
The network has 7,980 nodes and 21,036 edges.

\textbf{ASTRO-PH} \cite{snapnets}: This is a collaboration network of authors of papers submitted to e-print arXiv during the period from January 1993 to April 2003.
The network has 18,772 nodes and 396,160 edges.


\textbf{PROTEIN-PROTEIN INTERACTIONS (PPI)} \cite{breitkreutz2008biogrid}: This is a network of biological interactions between proteins in humans.
This network has 3,890 nodes and 38,739 edges.

\subsection{Evaluation Metrics}
To evaluate the performance of embedding methods on graph reconstruction and link prediction, we use Precision at $k$ ($Pr@k$) and $Mean Average Precision (MAP)$ as our metrics. 
For node classification, we use \textit{micro-F1} and \textit{macro-F1}. These metrics are defined as follows:

\textbf{\textit{Pr@k}} is the fraction of correct predictions in top $k$ predictions. It is defined as 
          $Pr@k = \frac{|E_{pred}(1:k) \cap E_{obs}|}{k}$,
    	where $E_{pred}(1:k)$ are the top $k$ predictions and $E_{obs}$ are the observed edges. 
    	For the task of graph reconstruction, $E_{obs} = E$ and for link prediction, $E_{obs}$ is the set of hidden edges.

\textbf{\textit{MAP}} estimates precision for every node and computes the average over all nodes, as follows:
        \begin{equation*}
        	MAP = \frac{\sum_i AP(i)}{|V|},
        \end{equation*}
        where $AP(i) = \frac{\sum_k Pr@k(i) \cdot \mathbb{I}\{E_{pred_i}(k) \in E_{obs_i}\}}{|\{k: E_{pred_i}(k) \in E_{obs_i}\}|}$, $Pr@k(i) = \frac{|E_{pred_i}(1:k) \cap E_{obs_i}|}{k}$, 
        and $E_{pred_i}$ and $E_{obs_i}$ are the predicted and observed edges for node $i$ respectively.

\textbf{\textit{macro-F1}}, in a multi-label classification task, is defined as the average $F1$ of all the labels, i.e.,
    	\begin{equation*}
    		macro-F1 = \frac{\sum_{l \in \mathcal{L}} F1(l)}{|\mathcal{L}|},
    	\end{equation*}
        where $F1(l)$ is the $F1$-score for label $l$.

\textbf{\textit{micro-F1}} calculates $F1$ globally by counting the total true positives, false negatives and false positives, giving equal weight to each instance. It is defined as follows:
    \begin{equation*}
    	micro-F1 = \frac{2*P*R}{P+R},
    \end{equation*}
    where
     	$P = \frac{\sum_{l \in \mathcal{L}} TP(l)}{\sum_{l \in \mathcal{L}} (TP(l) + FP(l))}$, and 
    	$R = \frac{\sum_{l \in \mathcal{L}} TP(l)}{\sum_{l \in \mathcal{L}} (TP(l) + FN(l))}$,
are precision (P) and recall (R) respectively, and 
$TP(l)$, $FP(l)$ and $FN(l)$ denote the number of true positives, false positives and false negatives respectively among the instances which are associated with the label $l$ either in the ground truth or the predictions.

\section{Experiments and Analysis}\label{sec:exp_analysis}
In this section, we evaluate and compare embedding methods on the for tasks presented above. For each task, we show the effect of number of embedding dimensions on the performance and compare hyper parameter sensitivity of the methods. Furthermore, we correlate the performance of embedding techniques on various tasks varying hyper parameters to test the notion of an ``all-good'' embedding which can perform well on all tasks.

\subsection{Graph Reconstruction}
\begin{figure*}[!ht]
	\centering
	\includegraphics[width=0.98\textwidth]{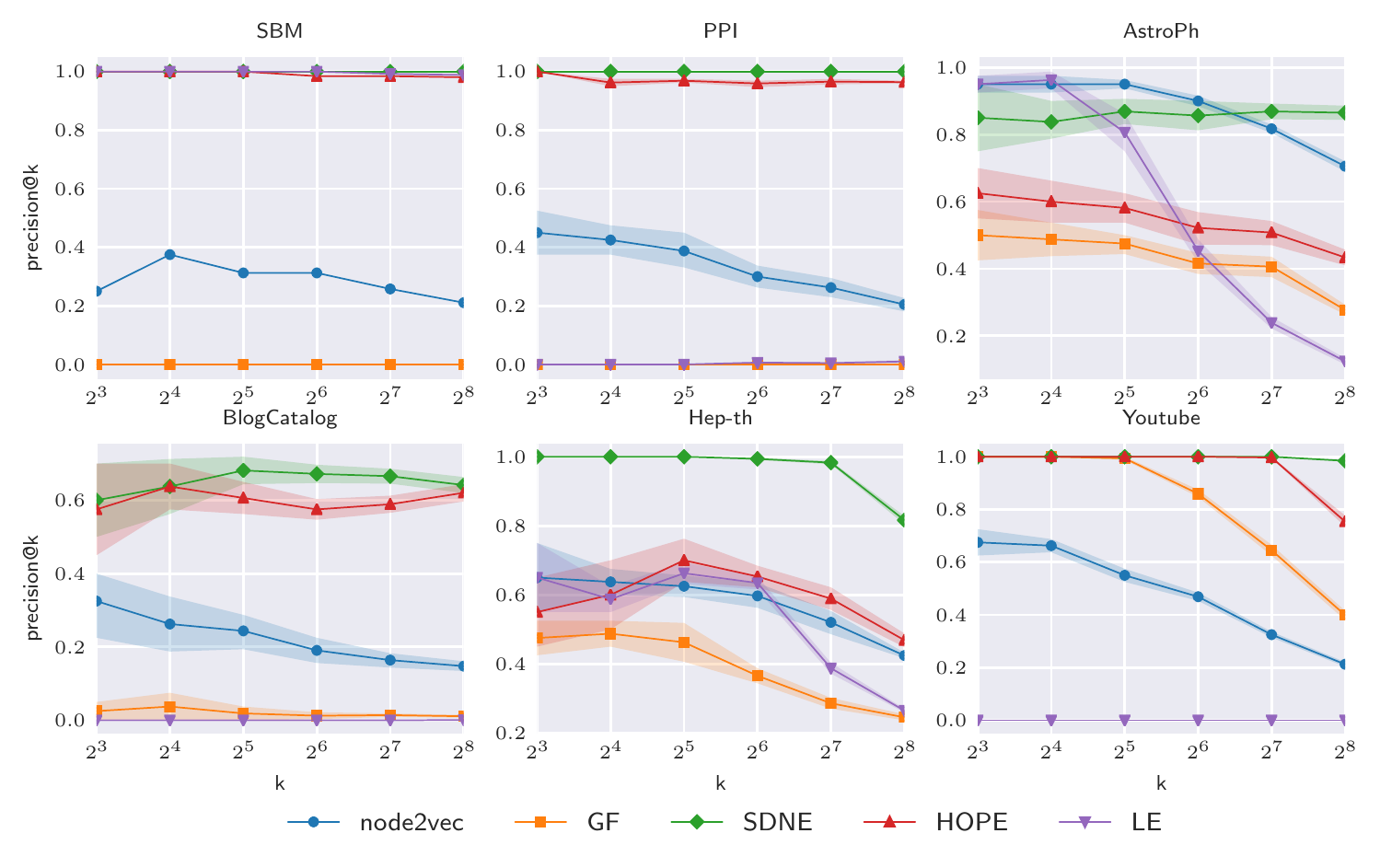}
	\hfil
	\caption{Precision@k of graph reconstruction for different data sets (dimension of embedding is 128).}
	\label{fig:p_at_k_128_gr}
\end{figure*}

\begin{figure*}[!ht]
	\centering
	\includegraphics[width=0.98\textwidth]{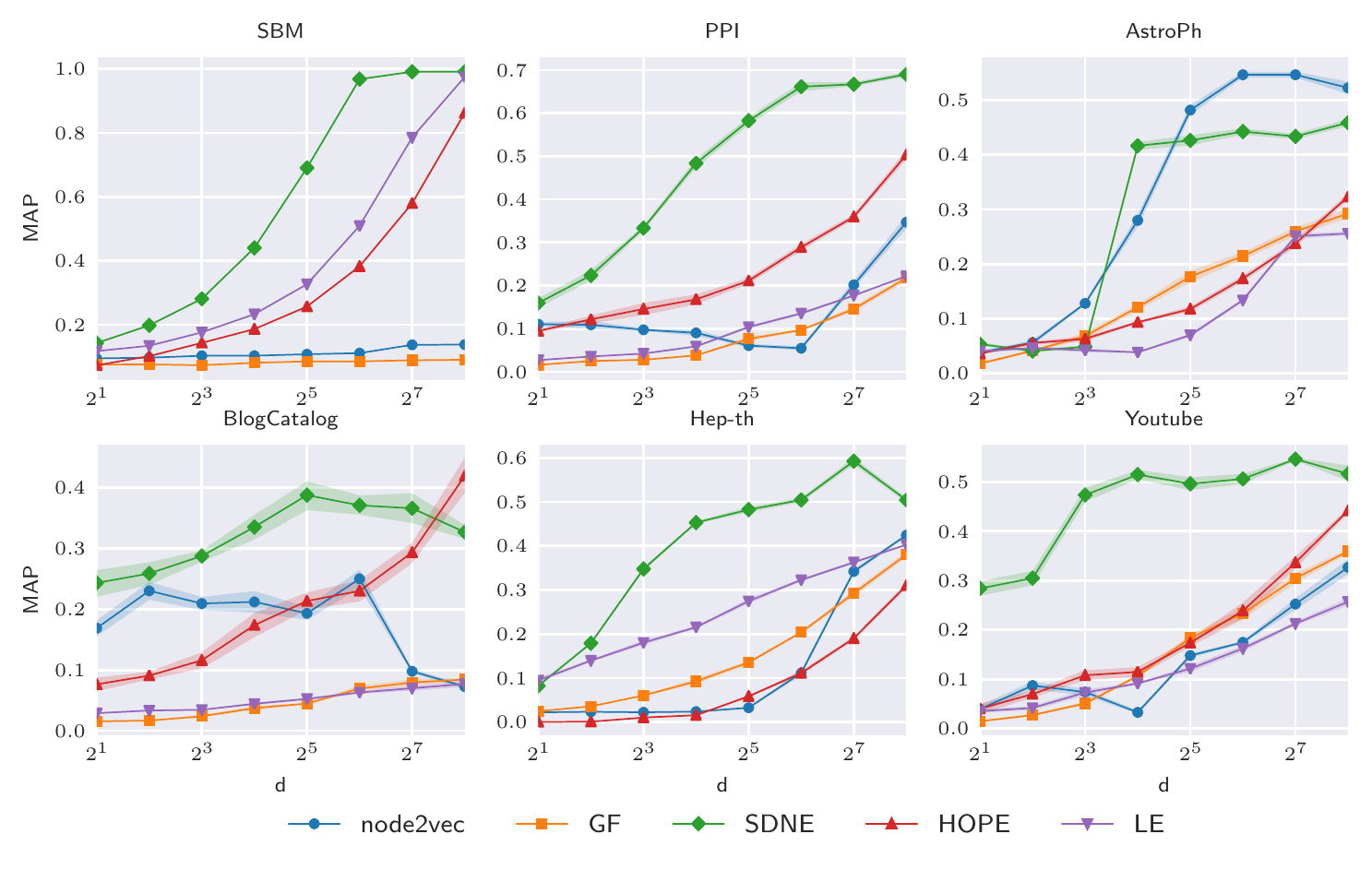}
	\hfil
	\caption{MAP of graph reconstruction for different data sets with varying dimensions.}
	\label{fig:map_gr}
\end{figure*}


Embeddings as a low-dimensional representation of the graph are expected to accurately reconstruct the graph. 
Note that reconstruction differs for different embedding techniques (refer to Section \ref{sec:approaches}).
For each method, we reconstruct the proximity of nodes and rank pair of nodes according to their proximity. 
Then we calculate the ratio of real links in top $k$ predictions as the reconstruction precision. 
As the number of possible node pairs ($N(N-1)$) can be very large for networks with a large number of nodes, we randomly sample 1024 nodes for evaluation.
We obtain 5 such samples for each dataset and calculate the mean and standard deviation of precision and MAP values for subgraph reconstruction.
To obtain optimal hyperparameters for each embedding method, we compare the mean MAP values for each hyperparameter.
We then re-run the experiments with the optimal hyperparameter on 5 random samples of 1024 nodes and report the results.

Figure \ref{fig:p_at_k_128_gr} illustrates the reconstruction precision obtained by 128-dimensional embeddings. We observe that although performance of methods is dataset dependent, embedding approaches which preserve higher order proximities in general outperform others. 
Exceptional performance of Laplacian Eigenmaps on SBM can be attributed to the lack of higher order structure in the data set. 
We also observe that SDNE consistently performs well on all data sets.
This can be attributed to its capability of learning complex structure from the network.
Embeddings learned by \textit{node2vec} have low reconstruction precision. 
This may be due to the highly non-linear dimensionality reduction yielding a non-linear manifold.
However, HOPE, which learns linear embeddings but preserves higher order proximity reconstructs the graph well without any additional parameters.

\textbf{Effect of dimension}. Figure \ref{fig:map_gr} illustrates the effect of dimension on the reconstruction error. 
With a couple of exceptions, as the number of dimensions increase, the MAP value increases. 
This is intuitive as higher number of dimensions are capable of storing more information. 
We also observe that SDNE is able to embed the graphs in 16-dimensional vector space with high precision although decoder parameters are required to obtain such precision.


\subsection{Visualization}
\begin{figure*}[!ht]
	\centering
	\subfloat[LLE]{\label{fig_m2} \includegraphics[width=0.32\textwidth]{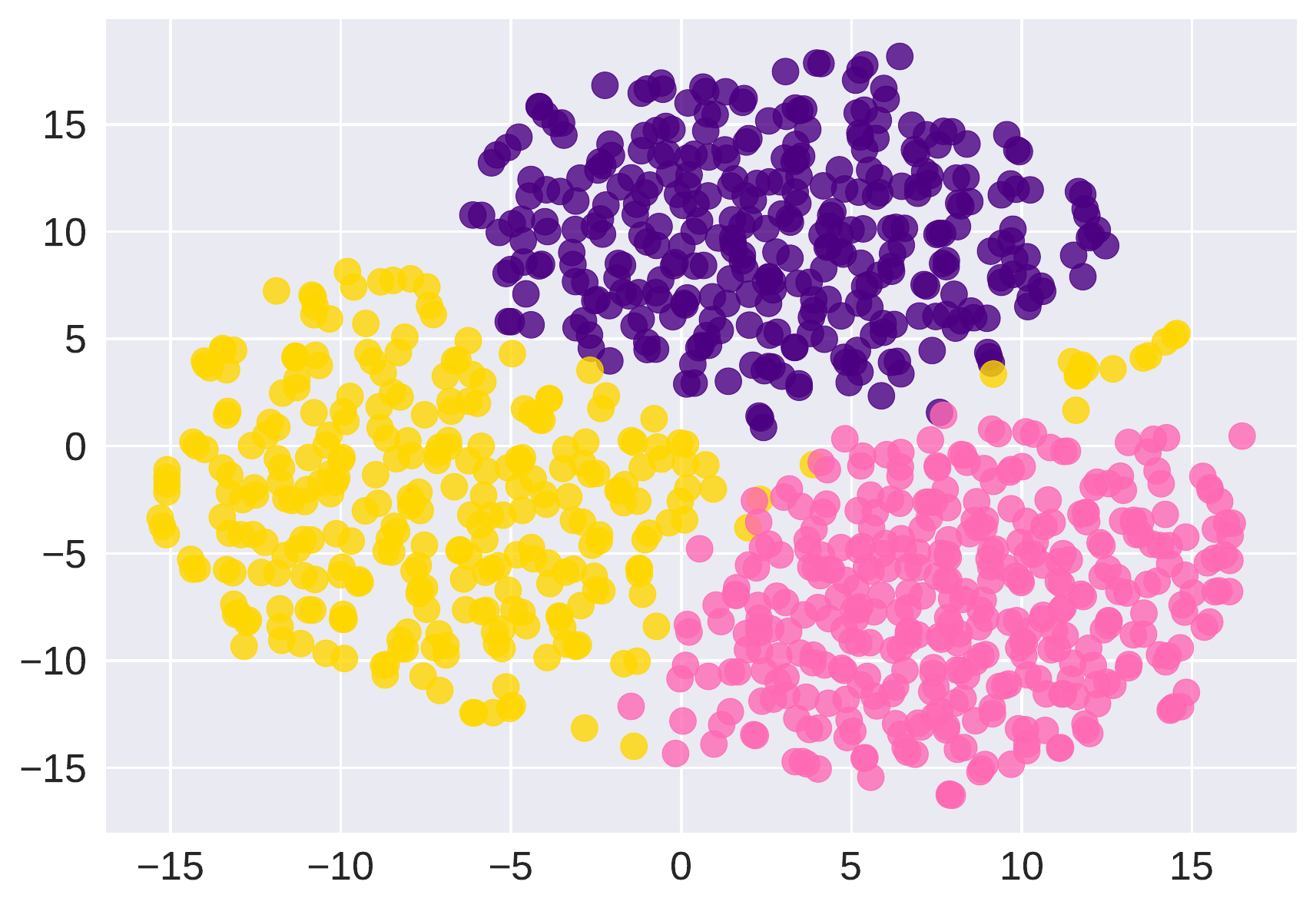}}
	\hfil
	\subfloat[GF]{\label{fig_m3} \includegraphics[width=0.32\textwidth]{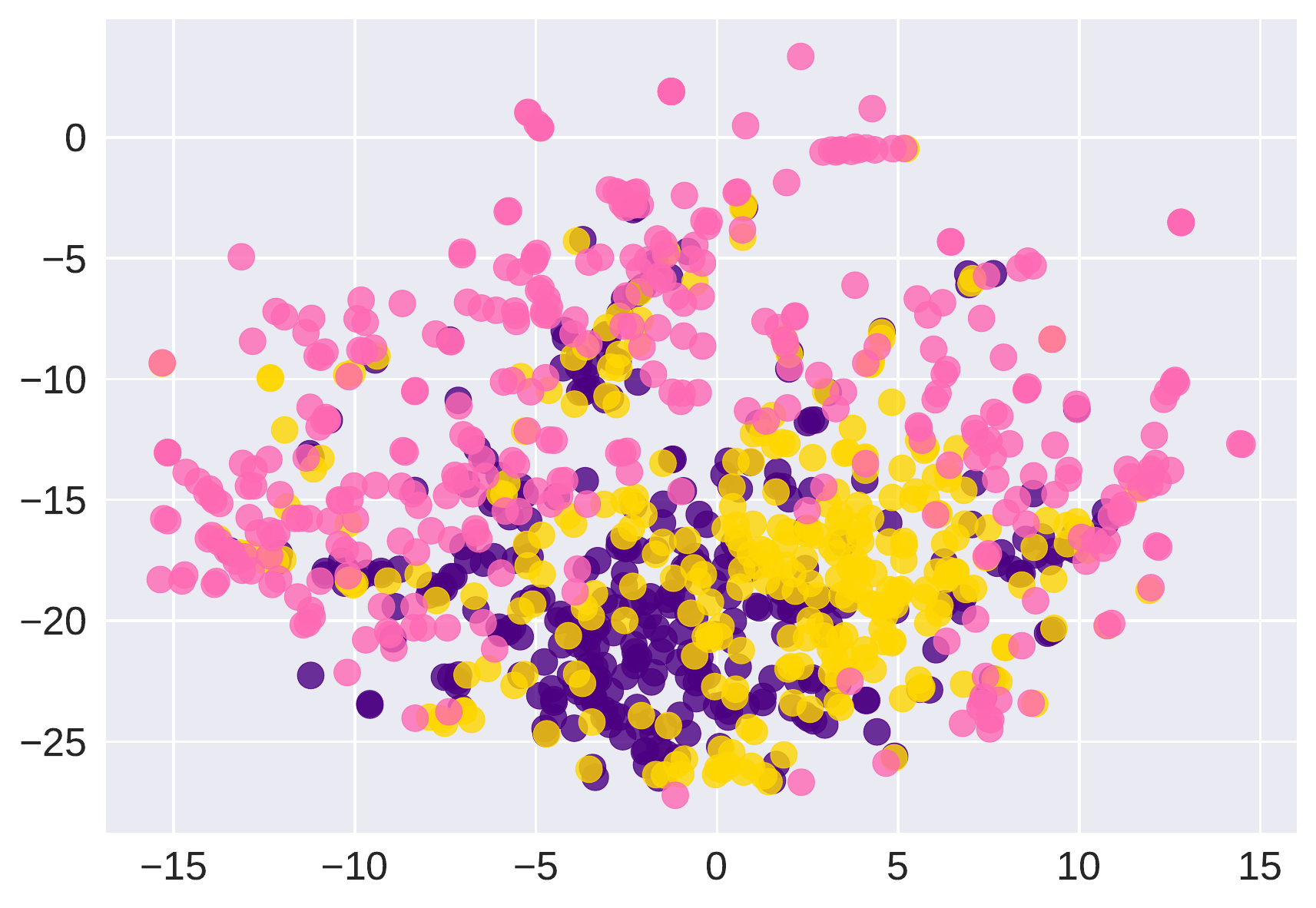}}
	\hfil
	\subfloat[\textit{node2vec}]{\label{fig_m4} \includegraphics[width=0.32\textwidth]{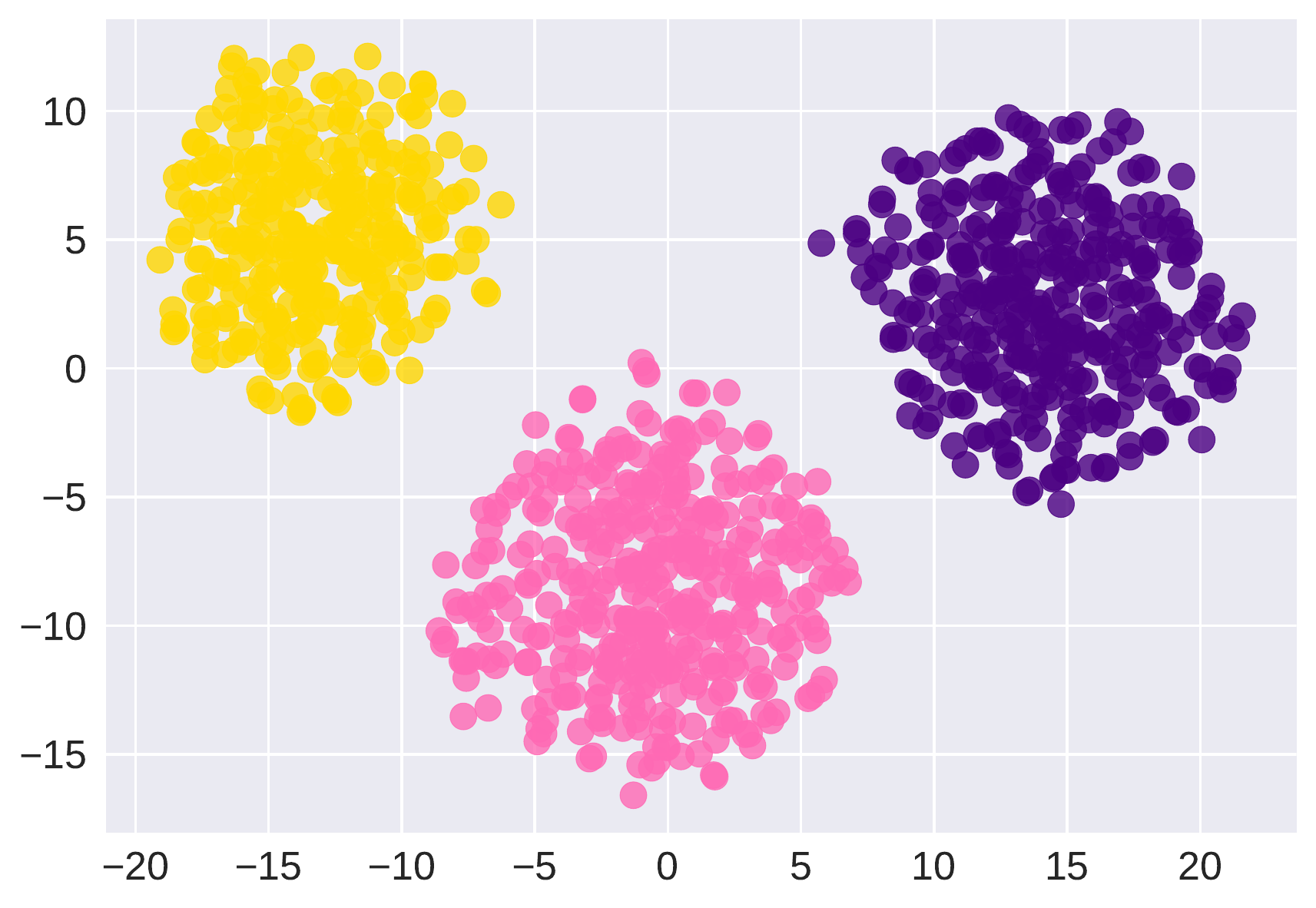}}
	\hfil
	\subfloat[HOPE]{\label{fig_m5} \includegraphics[width=0.32\textwidth]{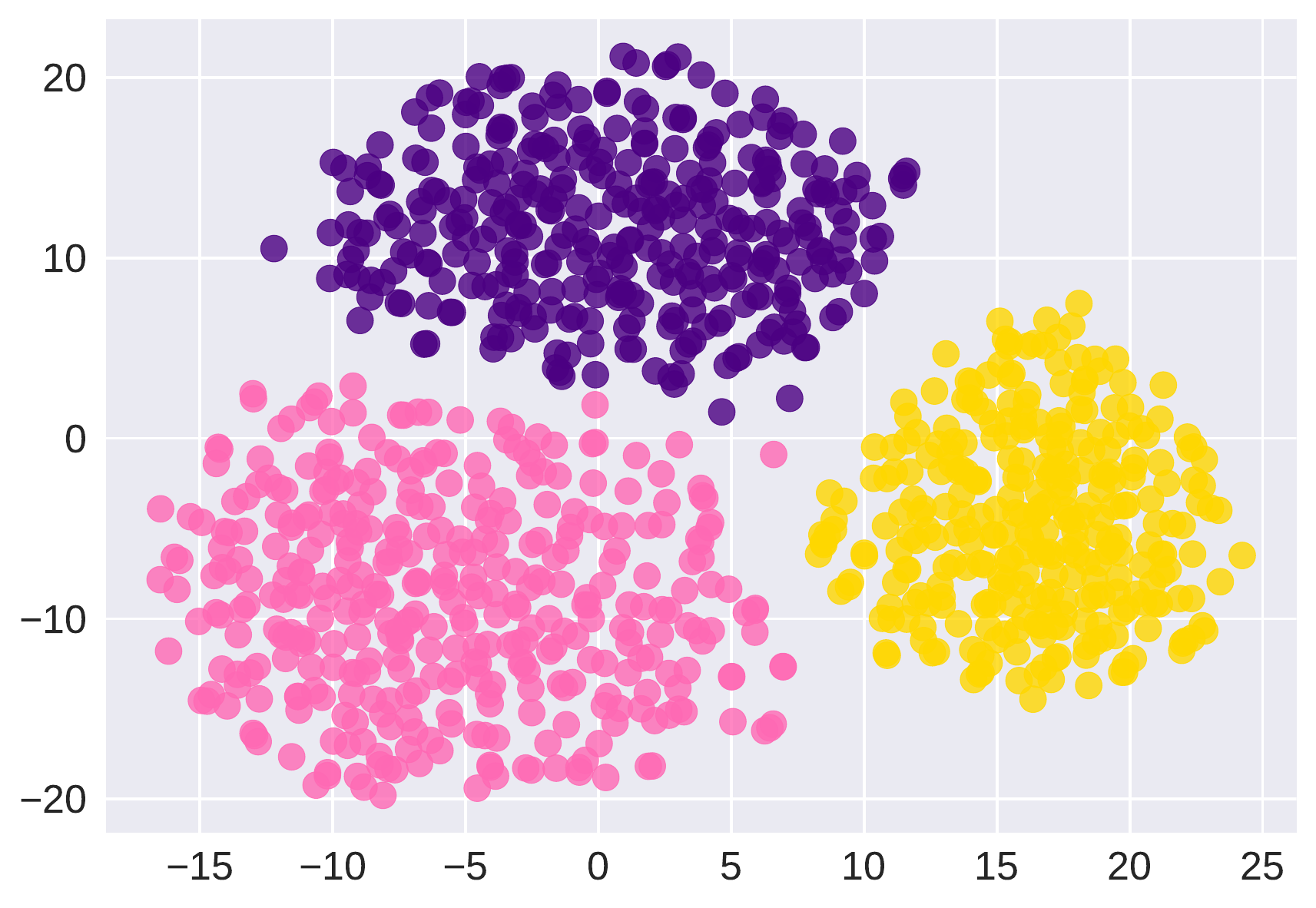}}
	\hfil
	\subfloat[SDNE]{\label{fig_m7} \includegraphics[width=0.32\textwidth]{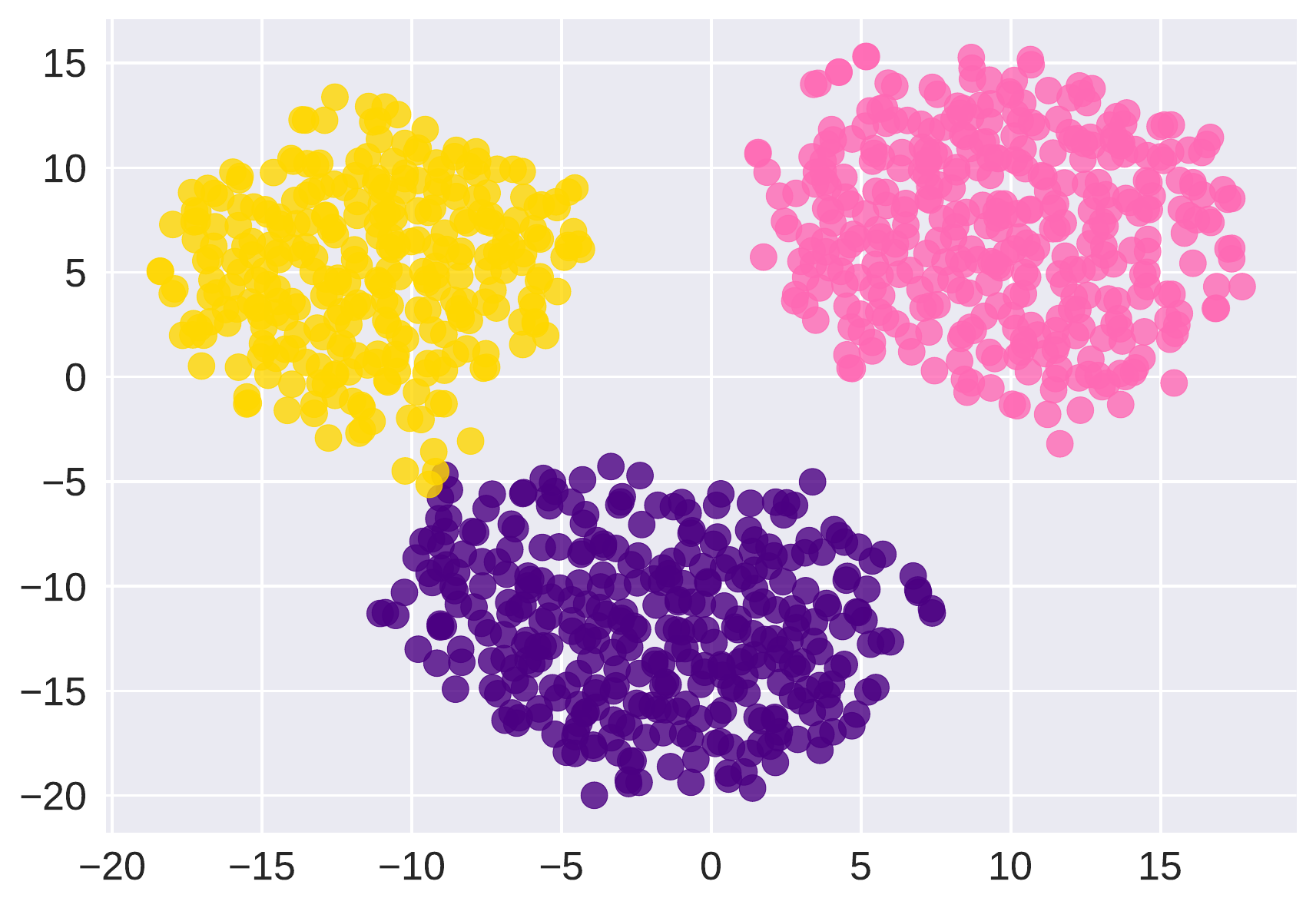}}
    \hfil
    \subfloat[LE]{\label{fig_m8} \includegraphics[width=0.32\textwidth]{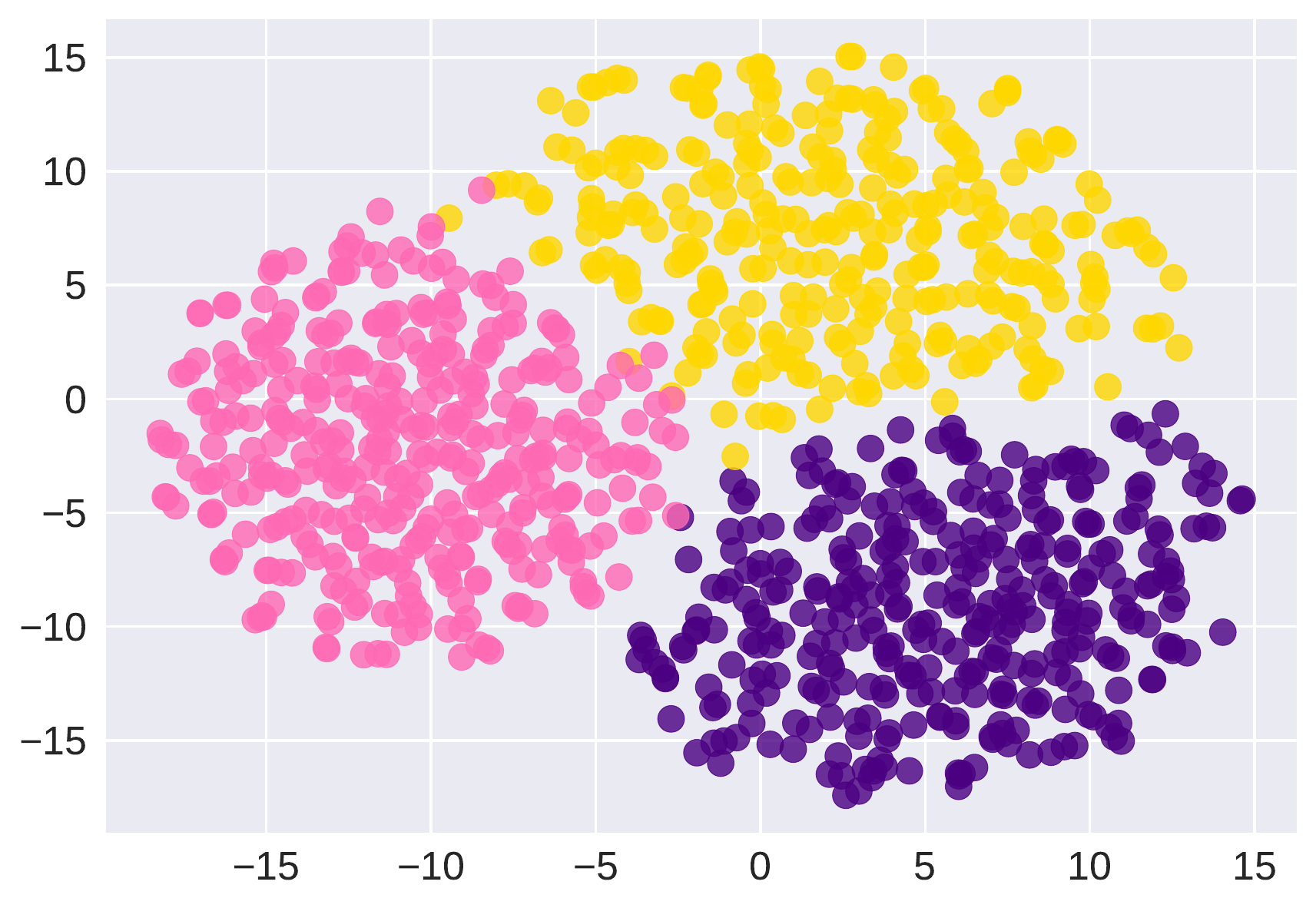}}
    \hfil
	\caption{Visualization of SBM using t-SNE (original dimension of embedding is 128). Each point corresponds to a node in the graph. Color of a node denotes its community.}
	\label{fig:viz_sbm}
\end{figure*}

\begin{figure*}[!ht]
	\centering
	\subfloat[LLE]{\label{fig_m2} \includegraphics[width=0.32\textwidth]{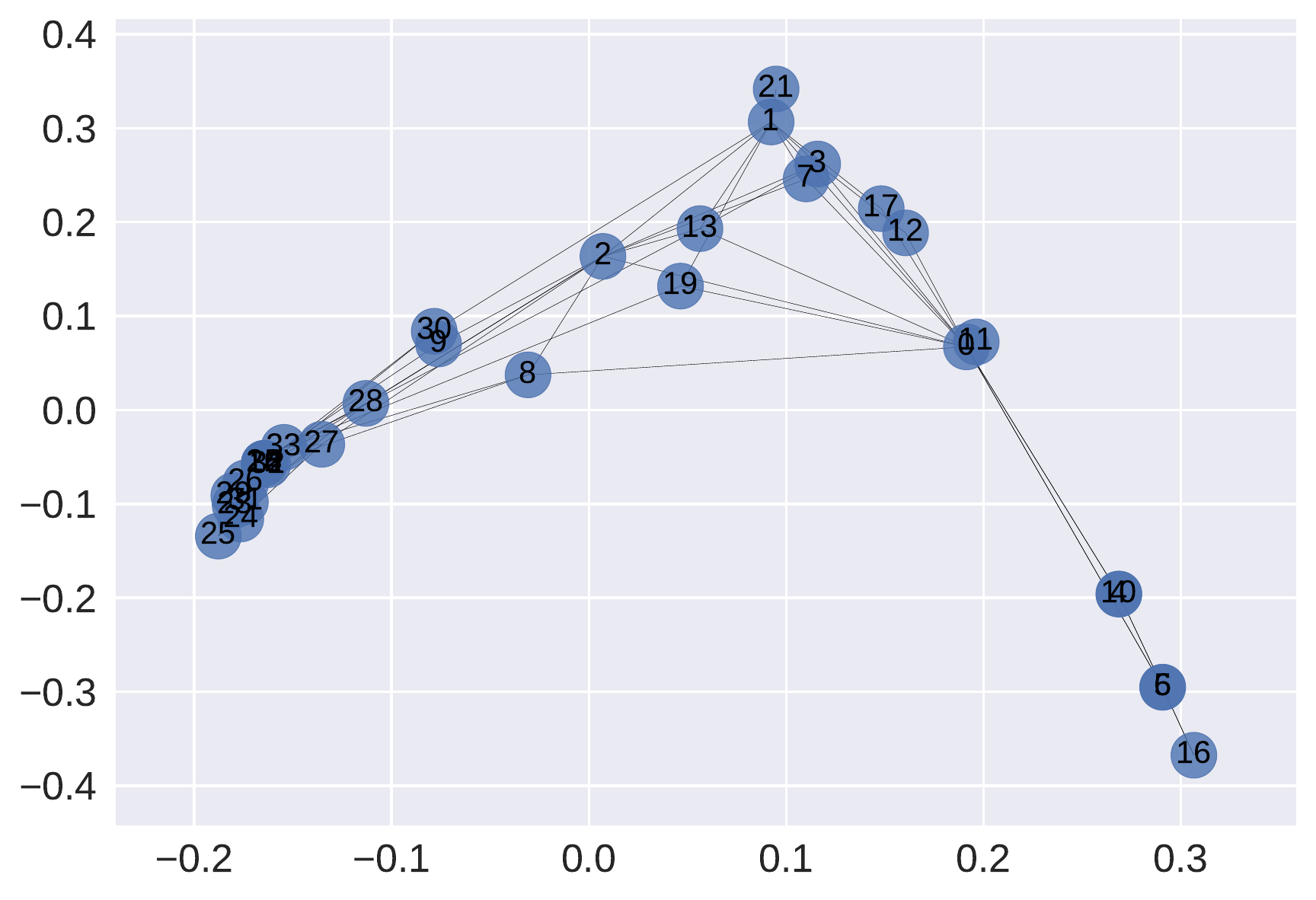}}
	\hfil
	\subfloat[GF]{\label{fig_m3} \includegraphics[width=0.32\textwidth]{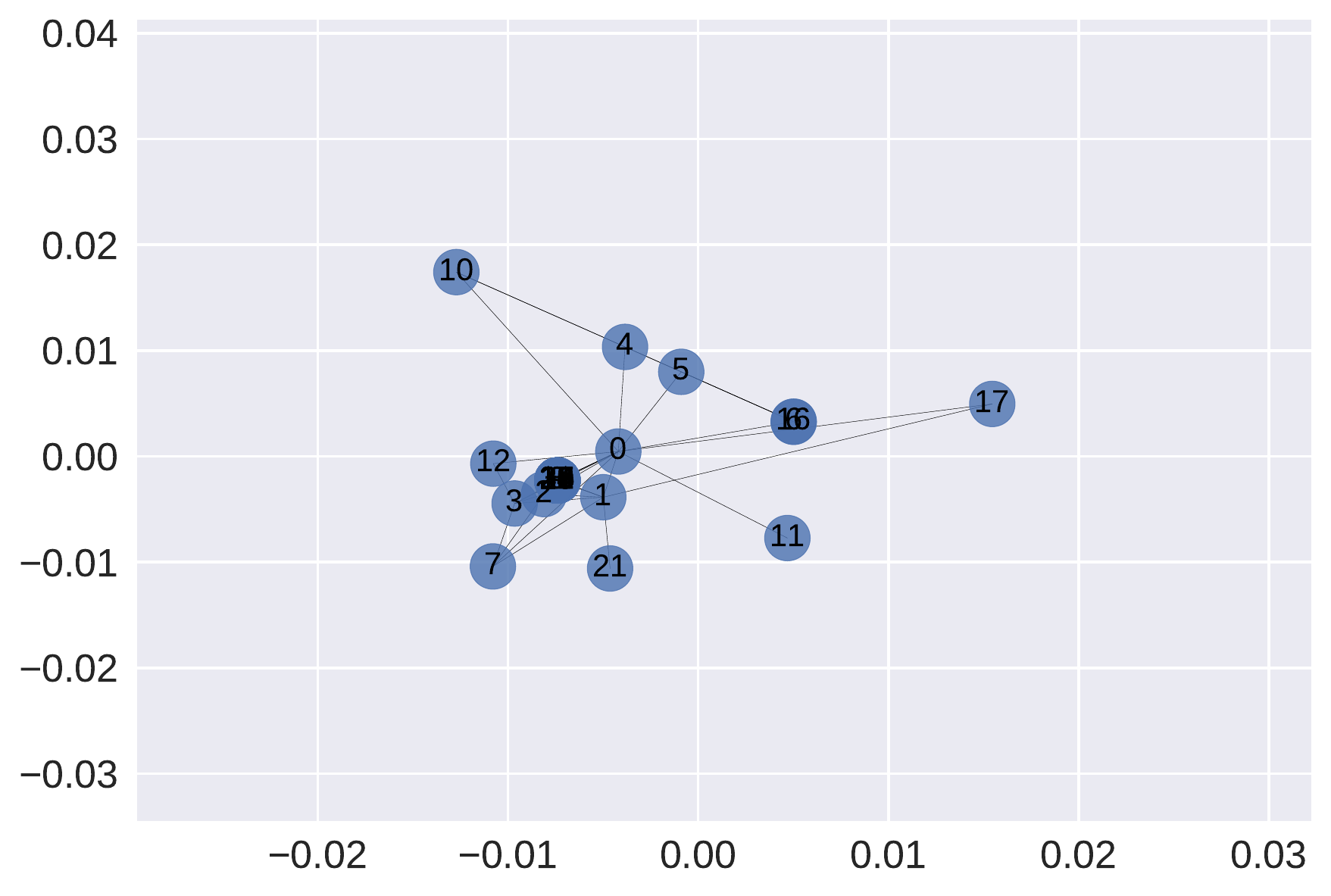}}
	\hfil
	\subfloat[\textit{node2vec}]{\label{fig_m4} \includegraphics[width=0.32\textwidth]{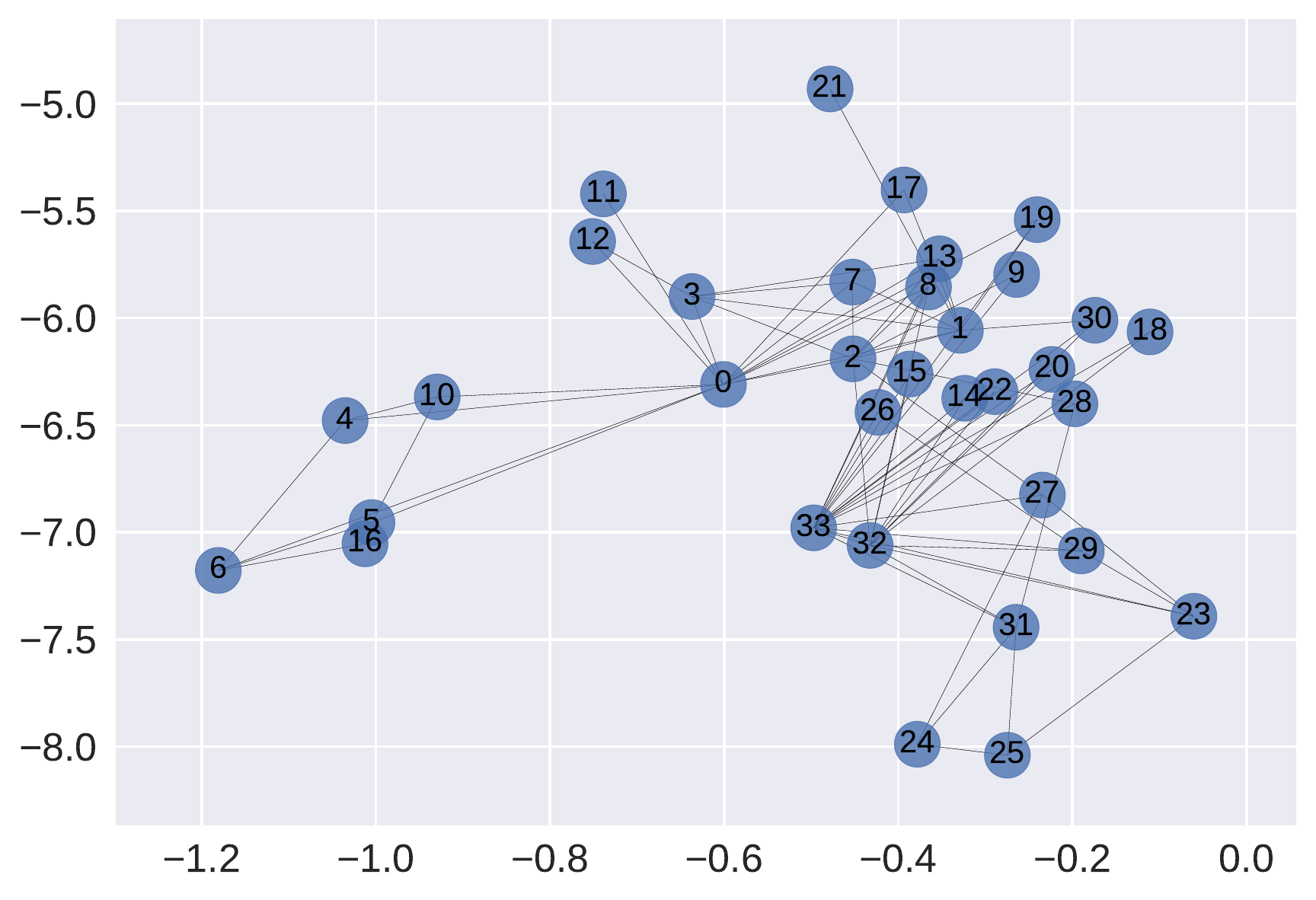}}
	\hfil
	\subfloat[HOPE]{\label{fig_m5} \includegraphics[width=0.32\textwidth]{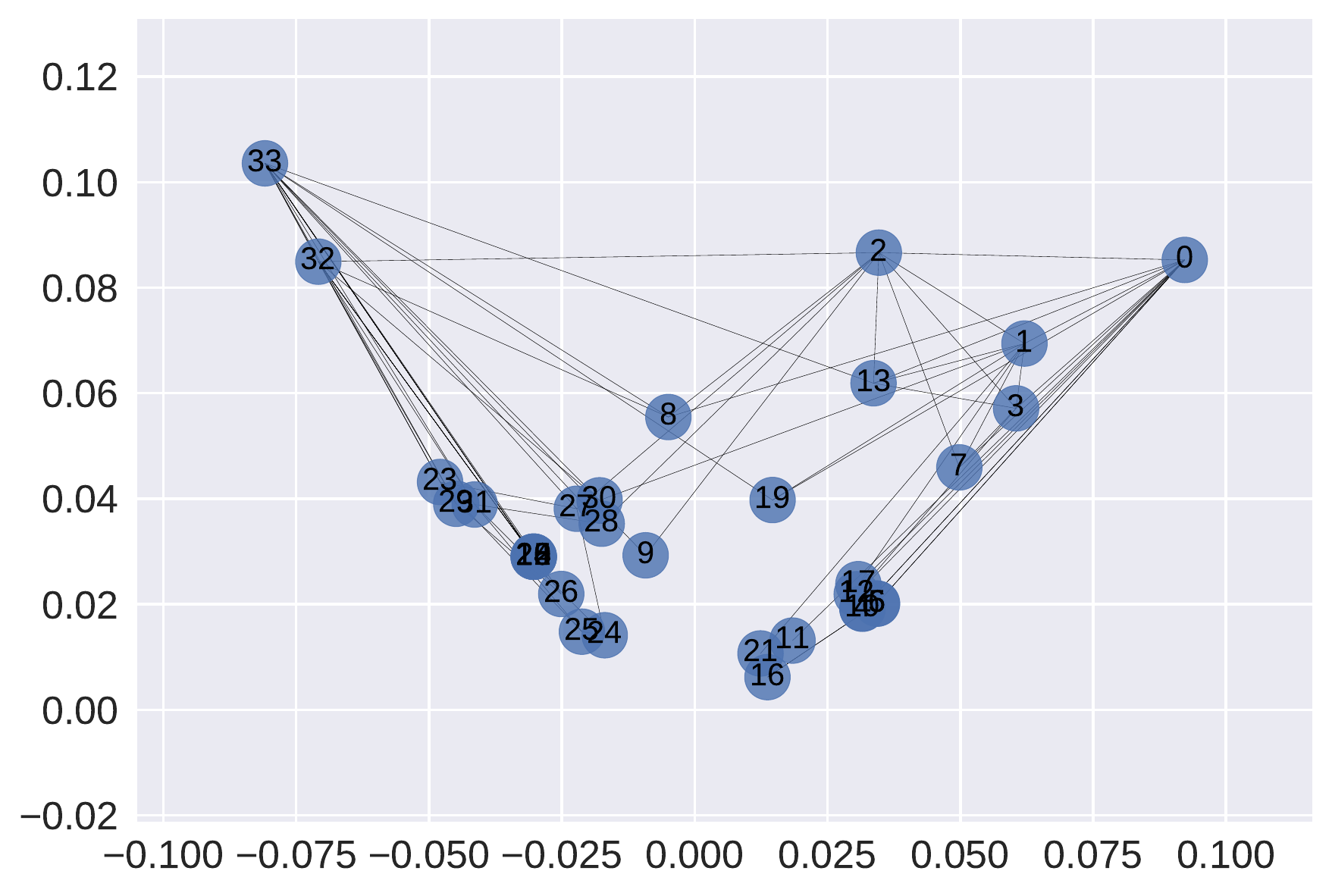}}
	\hfil
	\subfloat[SDNE]{\label{fig_m7} \includegraphics[width=0.32\textwidth]{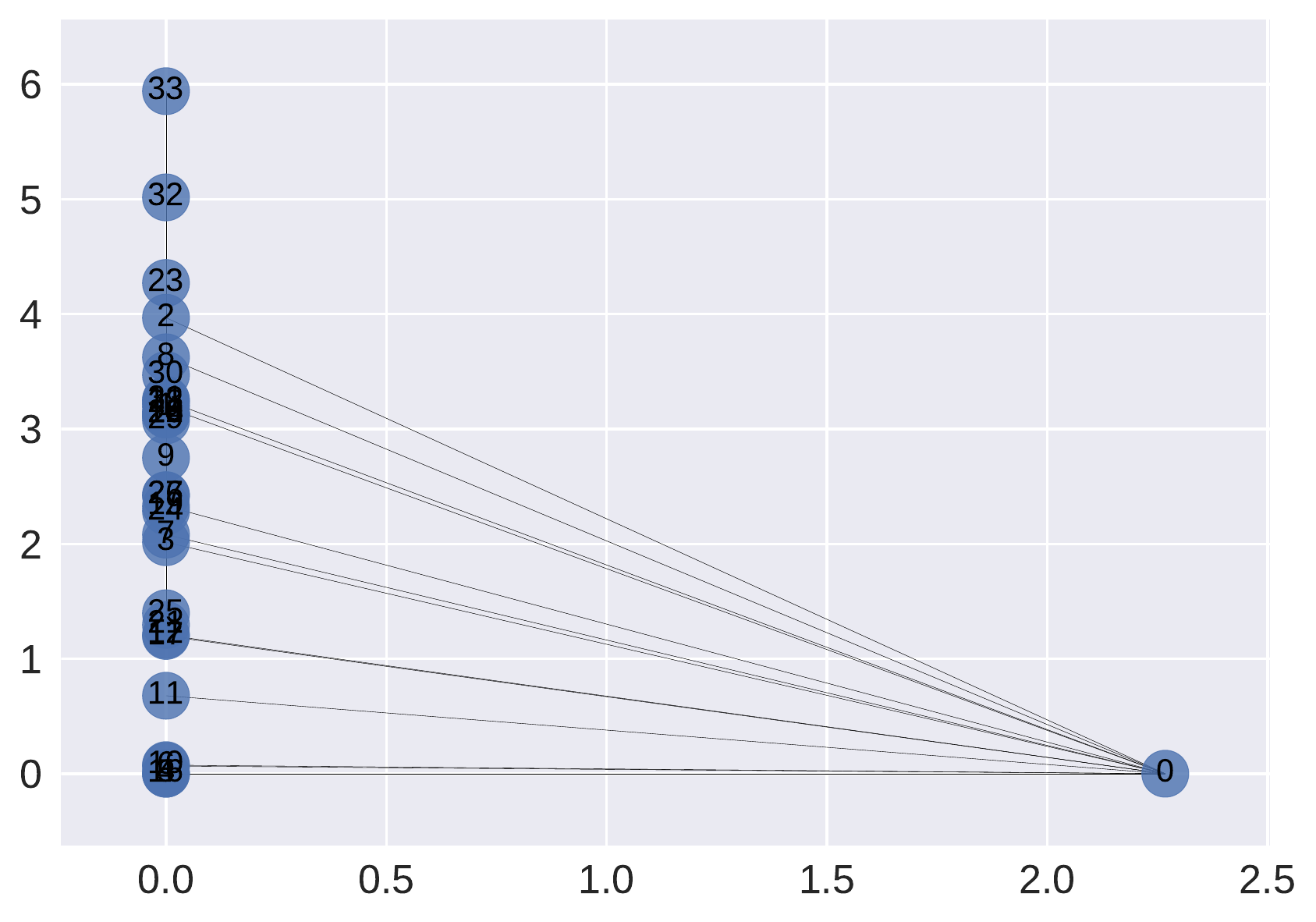}}
    \hfil
    \subfloat[LE]{\label{fig_m8} \includegraphics[width=0.32\textwidth]{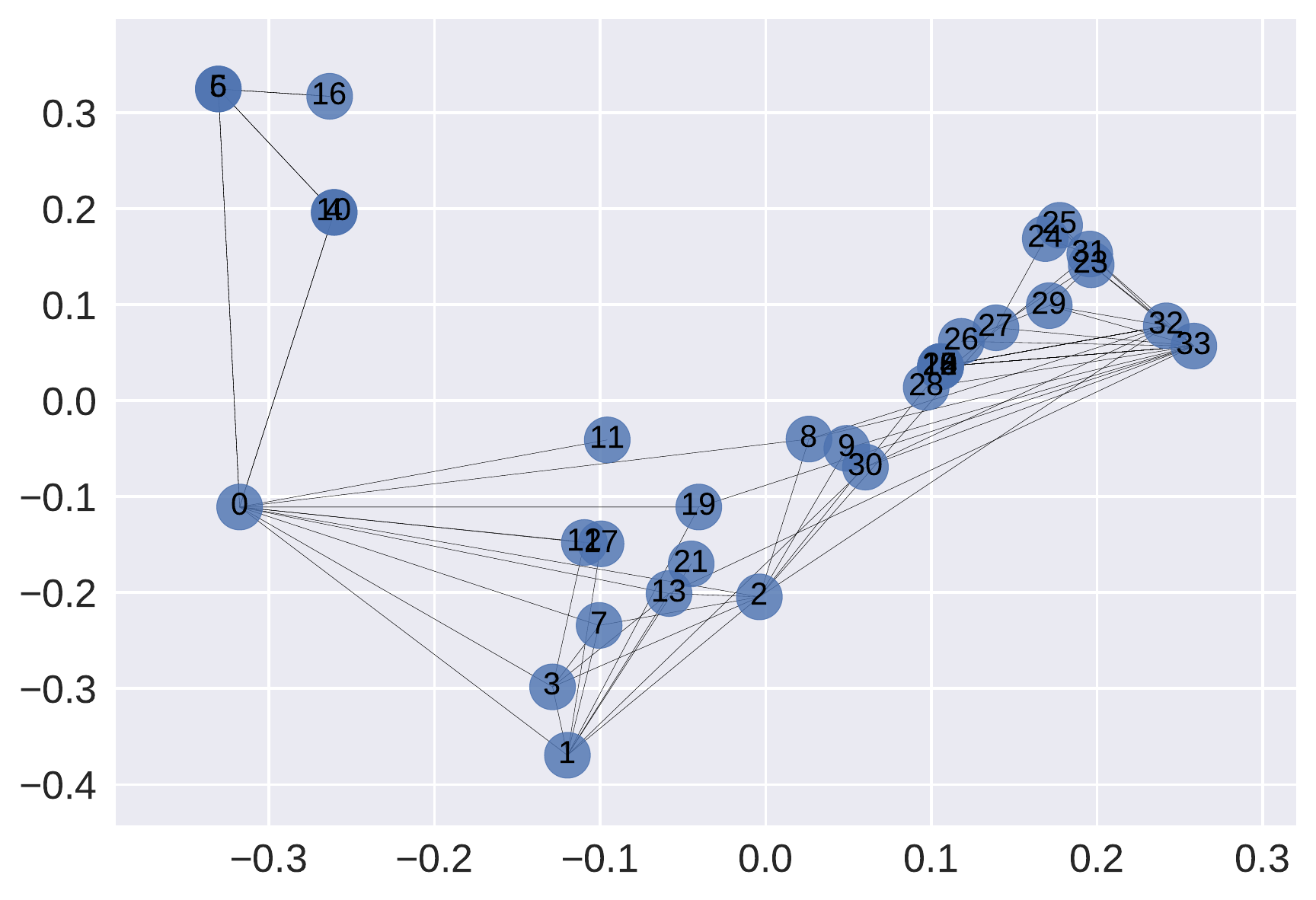}}
    \hfil
	\caption{Visualization of Karate club graph. Each point corresponds to a node in the graph.}
	\label{fig:viz_karate}
\end{figure*}

Since embedding is a low-dimensional vector representation of nodes in the graph, it allows us to visualize the nodes to understand the network topology. 
As different embedding methods preserve different structures in the network, their ability and interpretation of node visualization differ.
For instance, embeddings learned by \textit{node2vec} with parameters set to prefer BFS random walk would cluster structurally equivalent nodes together. 
On the other hand, methods which directly preserve $k$-hop distances between nodes (GF, LE and LLE with $k=1$ and HOPE and SDNE with $k>1$) cluster neighboring nodes together.
We compare the ability of different methods to visualize nodes on SBM and Karate graph.
For SBM, following \cite{Wang2016}, we learn a 128-dimensional embedding for each method and input it to t-SNE \cite{maaten2008visualizing} to reduce the dimensionality to 2 and visualize nodes in a 2-dimensional space.

Visualization of SBM is show in Figure \ref{fig:viz_sbm}.
As we know the underlying community structure, we use the community label to color the nodes.
We observe that embeddings generated by HOPE and SDNE which preserve higher order proximities well separate the communities although as the data is well structured LE, GF and LLE are able to capture community structure to some extent.

We visualize Karate graph (see Figure \ref{fig:viz_karate}) to illustrate the properties preserved by embedding methods.
LLE and LE ((a) and (f)) attempt to preserve the community structure of the graph and cluster nodes with high intra-cluster edges together.
GF ((b)) embeds communities very closely and keeps leaf nodes far away from other nodes. 
In (d), we observe that HOPE embeds nodes 16 and 21, whose Katz similarity in the original graph is very low (0.0006), farthest apart (considering dot product similarity).
\textit{node2vec} and SDNE ((c) and (e)) preserve a mix of community structure and structural property of the nodes.
Nodes 32 and 33, which are both high degree hubs and central in their communities, are embedded together and away from low degree nodes.
Also, they are closer to nodes which belong to their communities.
SDNE embeds node 0, which acts a bridge between communities, far away from other nodes.
Note that, unlike for other methods, it does not imply that node 0 is disconnected from the rest of the nodes.
The implication here is that SDNE identifies node 0 as a separate type of node and encodes its connection to other nodes in encoder and decoder.
The ability of deep autoencoders to identify important nodes in the network has not been studied but given this observation we believe this direction can be promising.

\subsection{Link Prediction}
\begin{figure*}[!ht]
	\centering
	\includegraphics[width=0.98\textwidth]{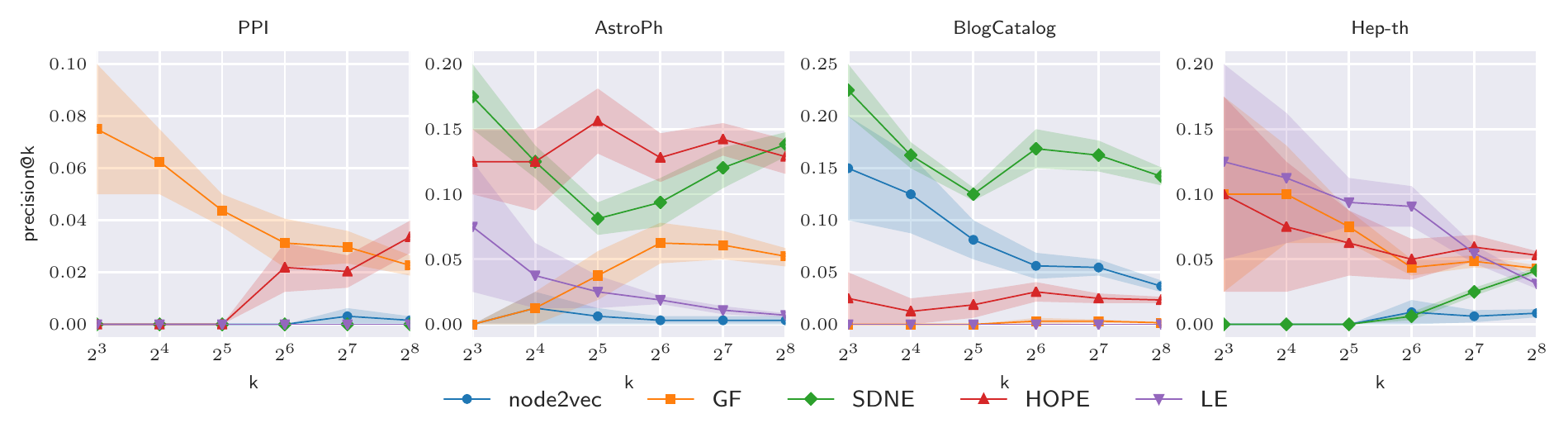}
	\hfil
	\caption{Precision@k of link prediction for different data sets (dimension of embedding is 128).}
	\label{fig:p_at_k_128_lp}
\end{figure*}

\begin{figure*}[!ht]
	\centering
	\includegraphics[width=0.98\textwidth]{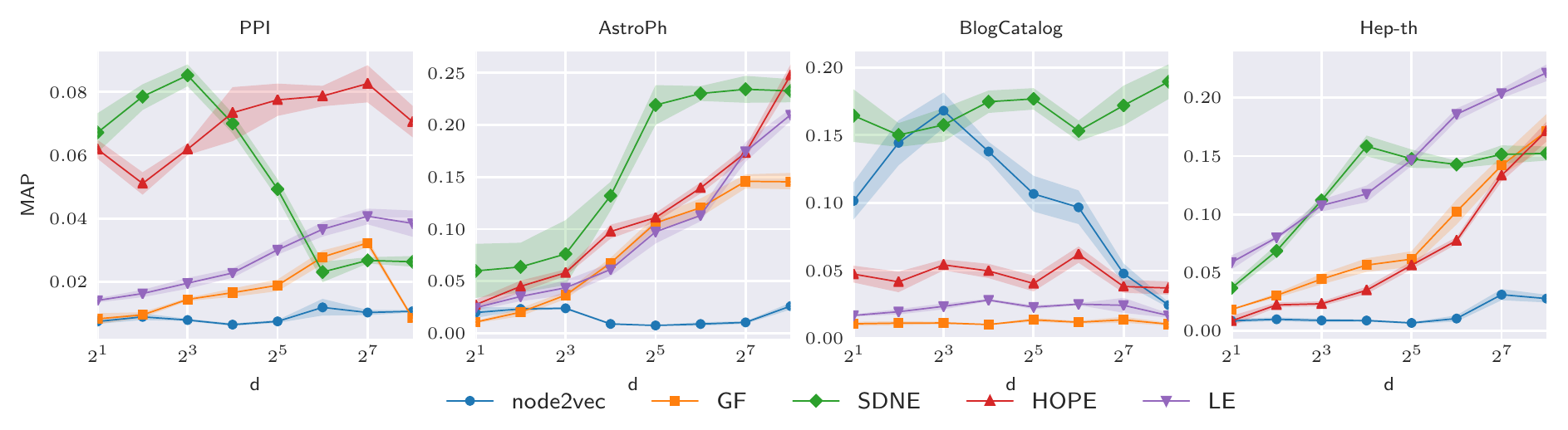}
	\hfil
	\caption{MAP of link prediction for different data sets with varying dimensions.}
	\label{fig:map_lp}
\end{figure*}

Another important application of graph embedding is predicting unobserved links in the graph. 
A good network representation should be able to capture the inherent structure of graph well enough to predict the likely but unobserved links. 
To test the performance of different embedding methods on this task, for each data set we randomly hide 20\% of the network edges. 
We learn the embedding using the rest of the 80\% edges and predict the most likely edges which are not observed in the training data from the learned embedding.
As with graph reconstruction, we generate 5 random subgraphs with 1024 nodes and test the predicted links against the held-out links in the subgraphs.
We perform this experiment for each hyperparameter and re-run it for optimal hyperparameters on a new random 80-20 link split.

Figure \ref{fig:p_at_k_128_lp} and \ref{fig:map_lp} show the $precision@k$ results for link prediction with 128-dimensional embeddings and MAP for each dimension respectively.
Here we can see that the performance of methods is highly data set dependent.
\textit{node2vec} achieves good performance on BlogCatalog but performs poorly on other data sets.
HOPE achieves good performance on all data sets which implies that preserving higher order proximities is conducive to predicting unobserved links.
Similarly, SDNE outperforms other methods with the exception on PPI for which the performance degrades drastically as embedding dimension increases above 8.

\textbf{Effect of dimension}. Figure \ref{fig:map_lp} illustrates the effect of embedding dimension on link prediction. 
We make two observations. 
Firstly, in PPI and BlogCatalog, unlike graph reconstruction performance does not improve as the number of dimensions increase. 
This may be because with more parameters the models overfit on the observed links and are unable to predict unobserved links.
Secondly, even on the same data set, relative performance of methods depends on the embedding dimension.
In PPI, HOPE outperforms other methods for higher dimensions,whereas embedding generated by SDNE achieves higher link prediction MAP for low dimensions.


\subsection{Node Classification}
\begin{figure*}[!ht]
	\centering
	\includegraphics[width=0.98\textwidth]{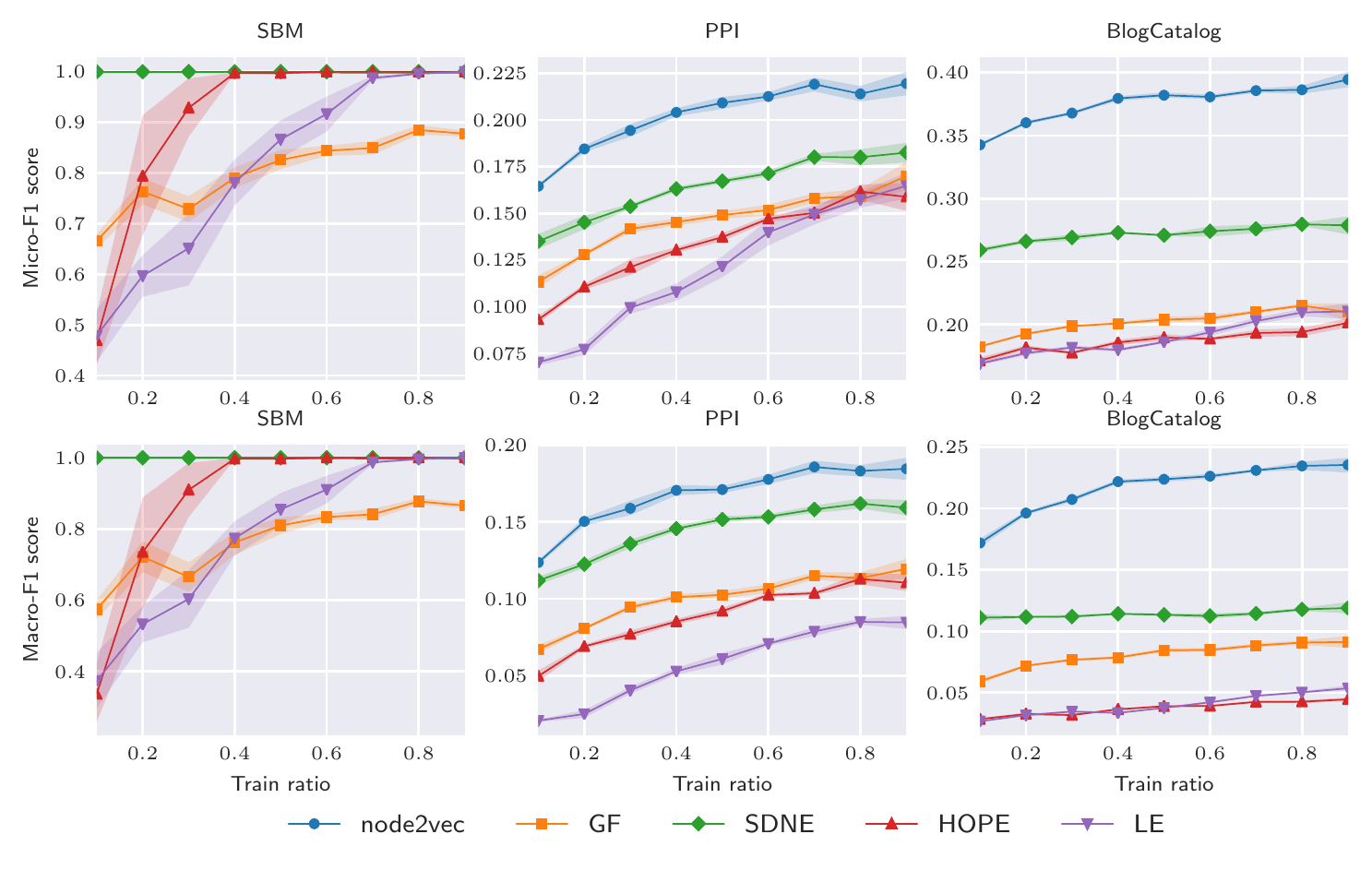}
	\hfil
	\caption{Micro-F1 and Macro-F1 of node classification for different data sets varying the train-test split ratio (dimension of embedding is 128).}
	\label{fig:nc_128}
\end{figure*}

\begin{figure*}[!ht]
	\centering
	\includegraphics[width=0.98\textwidth]{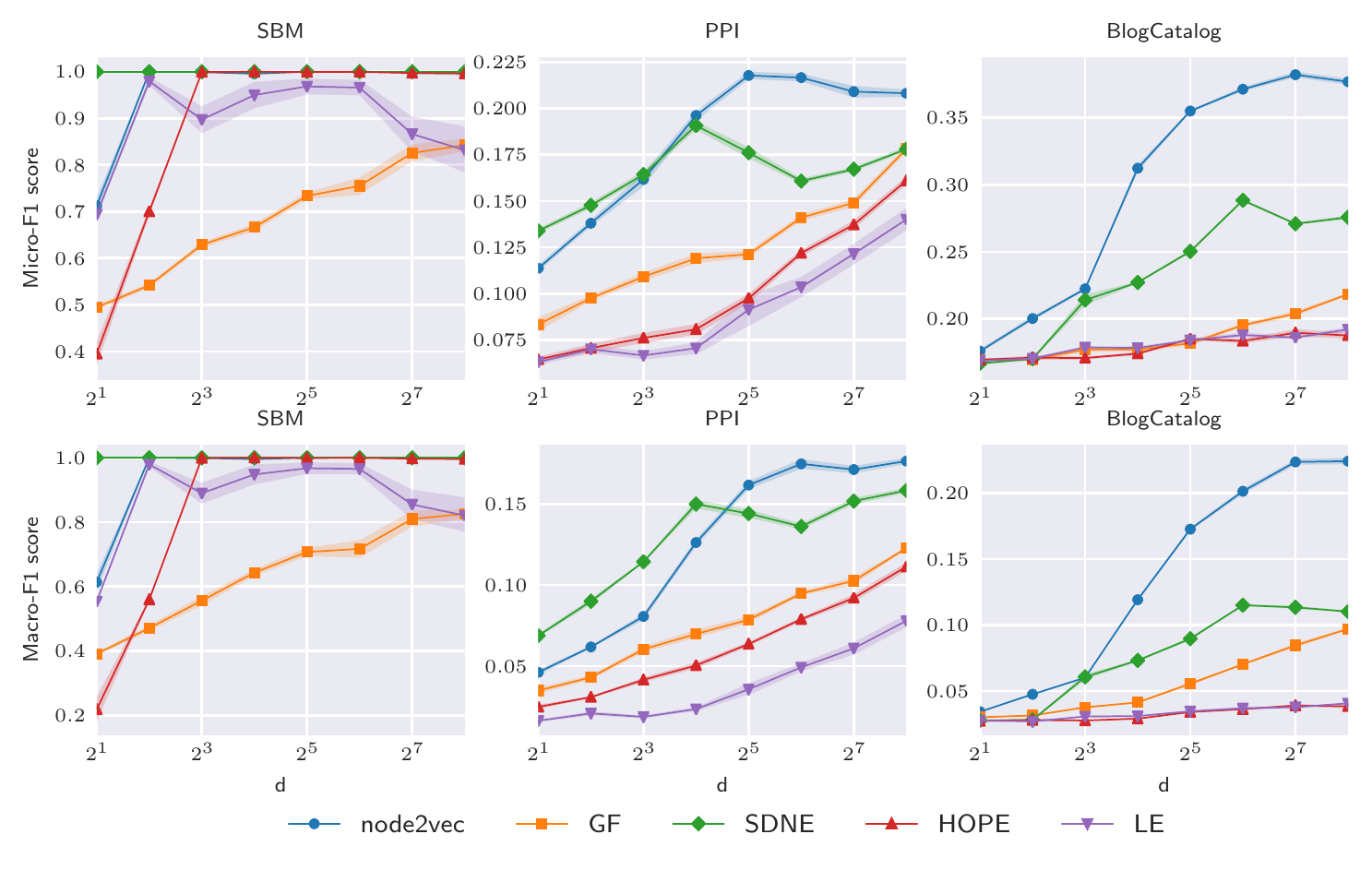}
	\hfil
	\caption{Micro-F1 and Macro-F1 of node classification for different data sets varying the number of dimensions. The train-test split is 50\%.}
	\label{fig:nc}
\end{figure*}

Predicting node labels using network topology is widely popular in network analysis and has variety of applications, including document classification and interest prediction.
A good network embedding should capture the network structure and hence be useful for node classification.
We compare the effectiveness of embedding methods on this task by using the generated embedding as node features to classify the nodes.
The node features are input to a one-vs-rest logistic regression using the LIBLINEAR  library.
For each data set, we randomly sample 10\% to 90\% of nodes as training data and evaluate the performance on the remaining nodes.
We perform this split 5 times and report the mean with confidence interval. 
For data sets with multiple labels per node, we assume that we know how many labels to predict.

Figure \ref{fig:nc_128} shows the results of our experiments.
We can see that \textit{node2vec} outperforms other methods on the task of node classification.
As mentioned earlier (\S\ref{sec:approaches}), \textit{node2vec} preserves homophily as well as structural equivalence between nodes.
Results suggest this can be useful in node classification: e.g., in BlogCatalog users may have similar interests, yet connect to others based on social ties rather than interests overlap.
Similarly, proteins in PPI may be related in functionality and interact with similar proteins but may not assist each other.
However, in SBM, other methods outperform \textit{node2vec} as labels reflect communities yet there is no structural equivalence between nodes.

\textbf{Effect of dimension}. Figure \ref{fig:nc} illustrates the effect of embedding dimensions on node classification.
As with link prediction, we observe that performance often saturates or deteriorates after certain number of dimensions.
This may suggest overfitting on the training data.
As SBM exhibits very structured communities, an 8-dimensional embedding suffices to predict the communities.
\textit{node2vec} achieves best performance on PPI and BlogCat with 128 dimensions.

\subsection{Hyperparameter Sensitivity}
\begin{figure*}[!ht]
	\centering
	\subfloat[Graph Reconstruction]{\label{fig_m2} \includegraphics[width=0.32\textwidth]{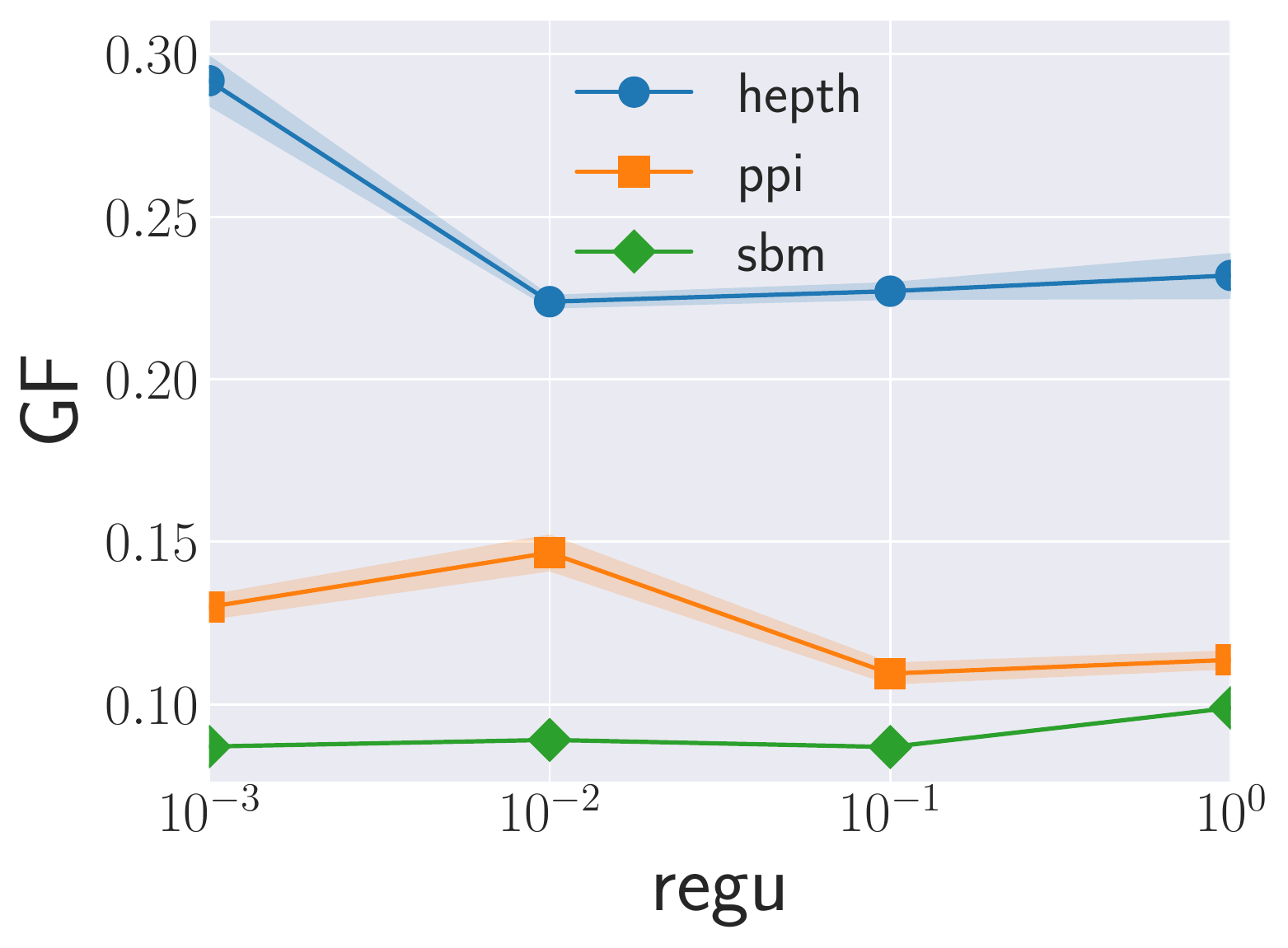}}
	\hfil
	\subfloat[Link Prediction]{\label{fig_m3} \includegraphics[width=0.32\textwidth]{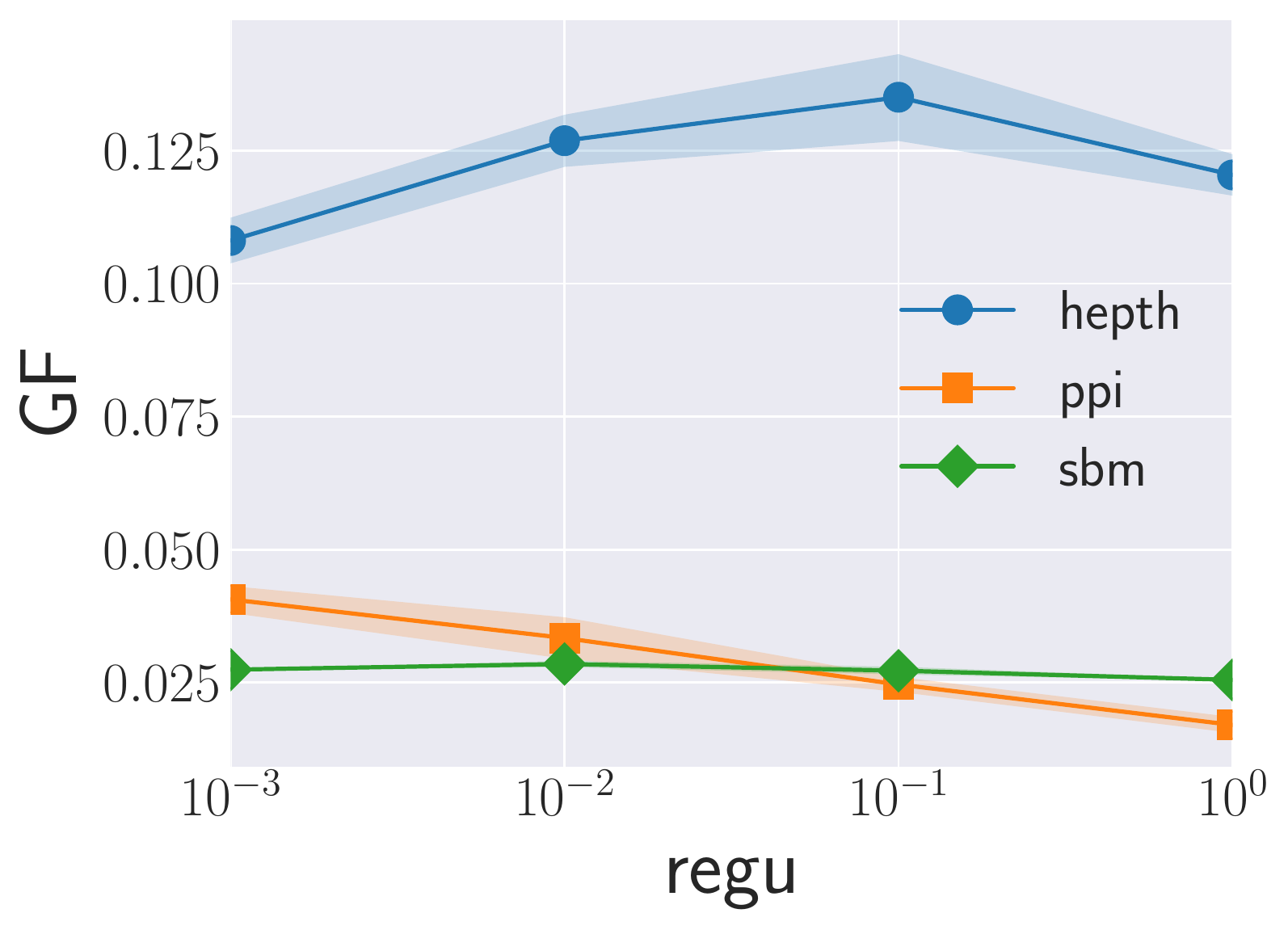}}
	\hfil
	\subfloat[Node Classification]{\label{fig_m4} \includegraphics[width=0.32\textwidth]{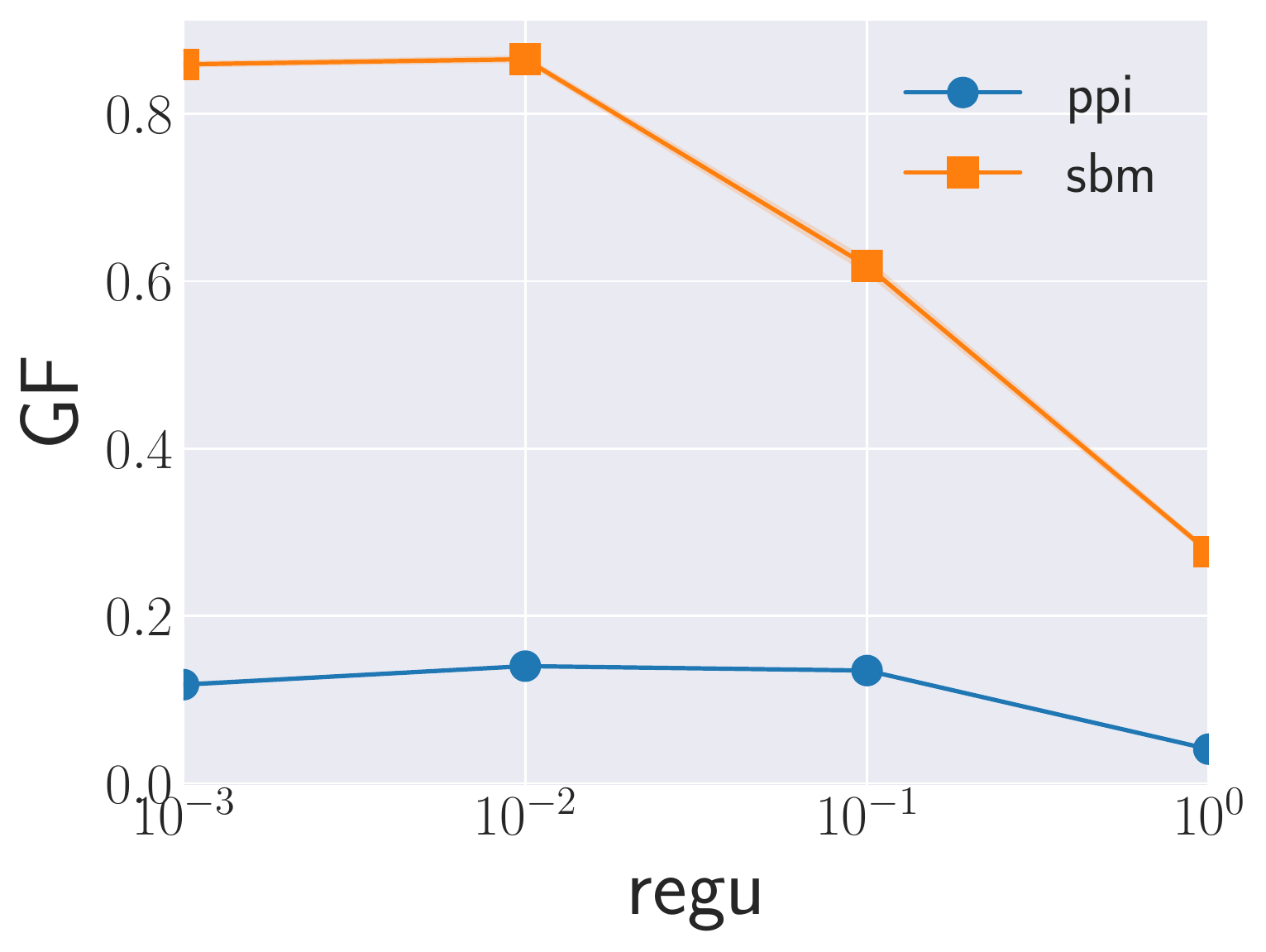}}
	\hfil
	\caption{Effect of regularization coefficient ($\alpha$) in Graph Factorization on various tasks.}
	\label{fig:gf_param}
\end{figure*}
\begin{figure*}[!ht]
	\centering
	\subfloat[Graph Reconstruction]{\label{fig_m2} \includegraphics[width=0.32\textwidth]{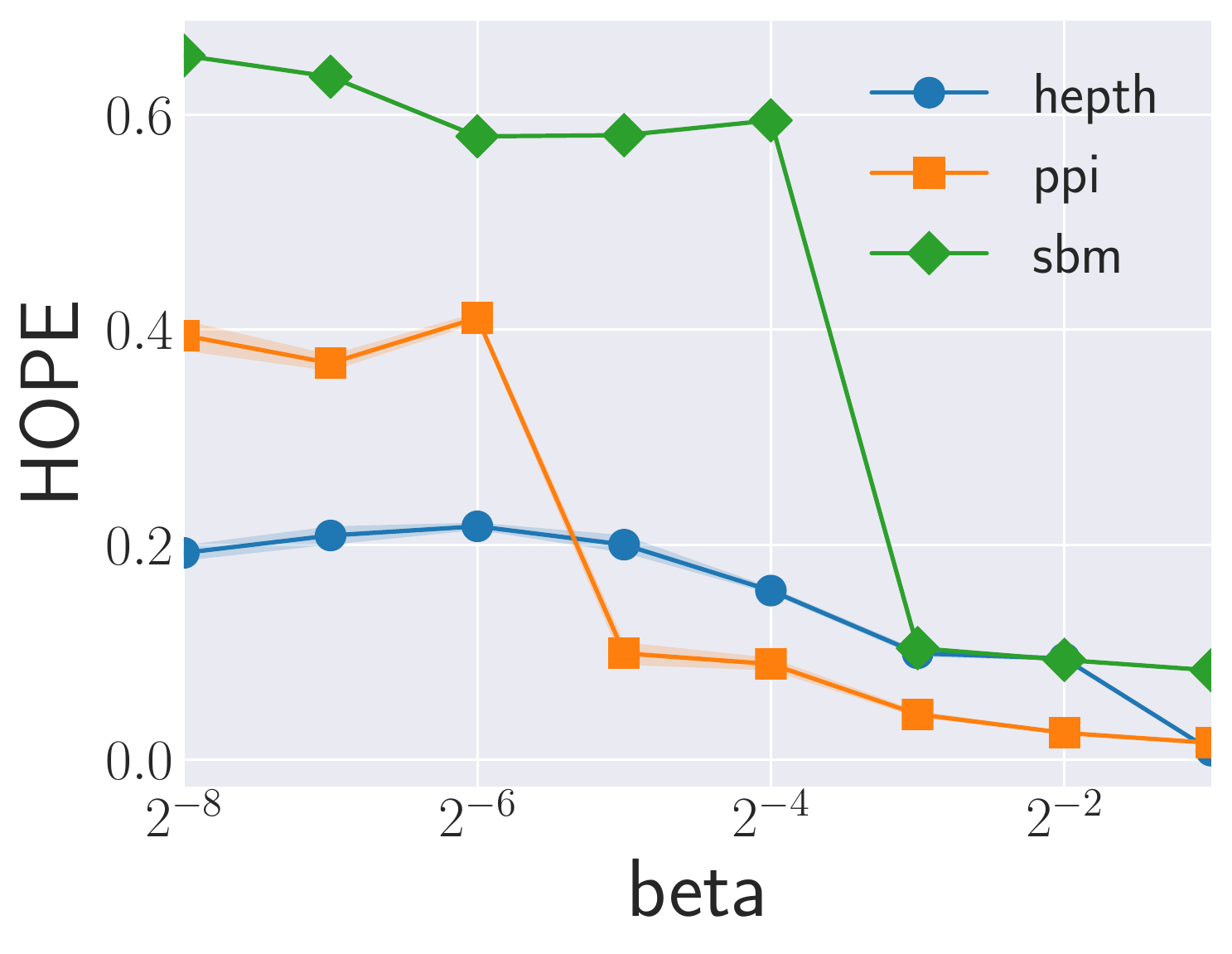}}
	\hfil
	\subfloat[Link Prediction]{\label{fig_m3} \includegraphics[width=0.32\textwidth]{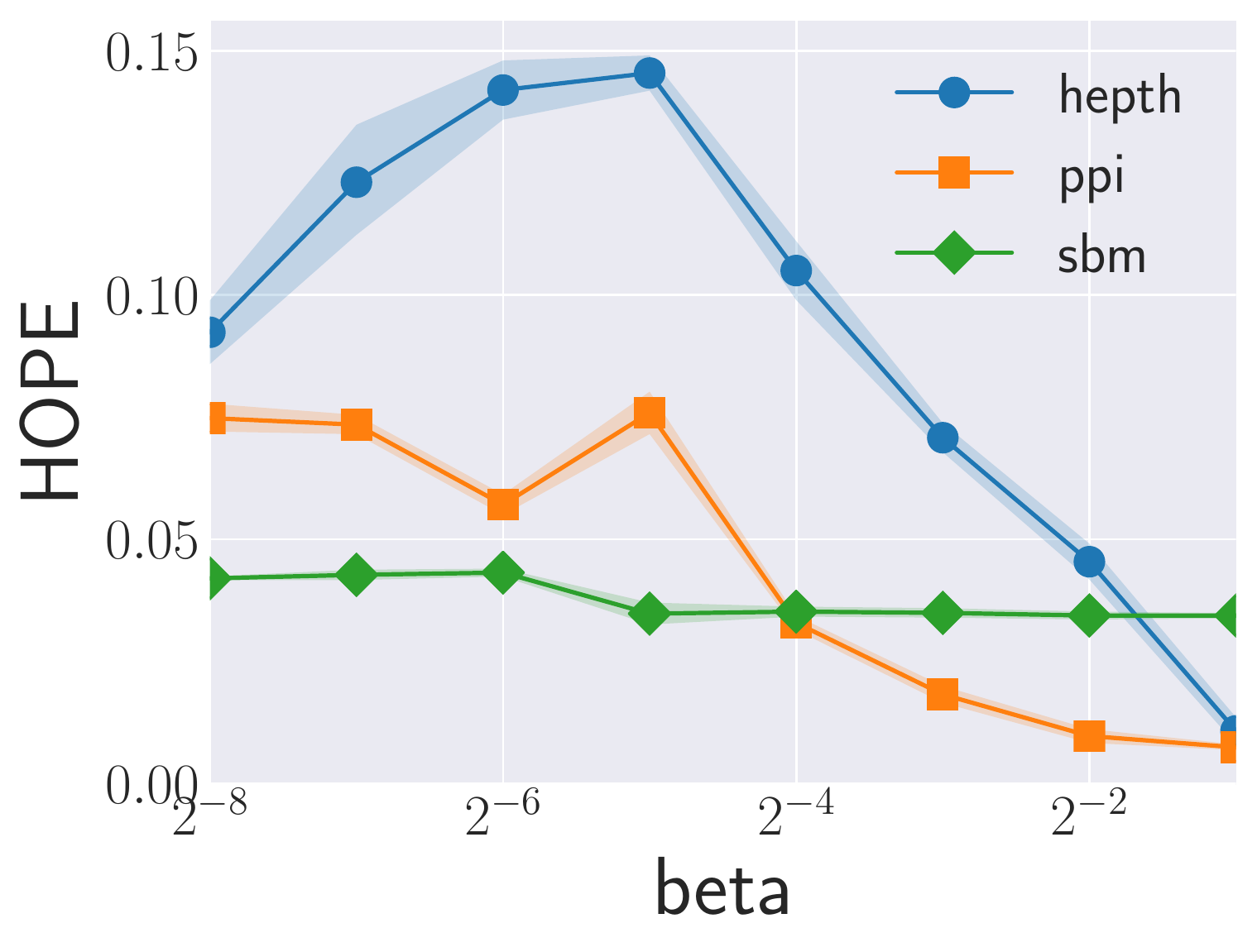}}
	\hfil
	\subfloat[Node Classification]{\label{fig_m4} \includegraphics[width=0.32\textwidth]{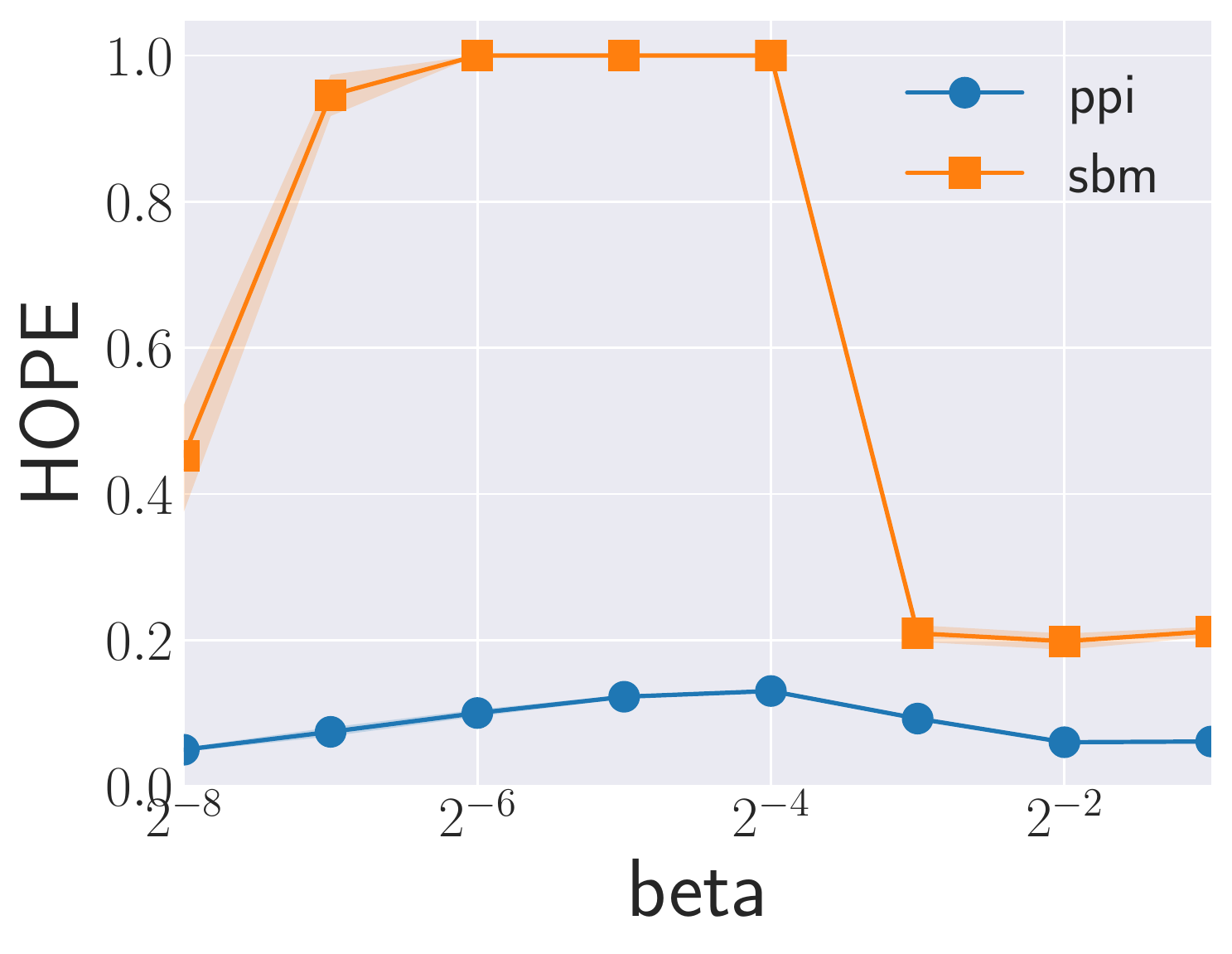}}
	\hfil
	\caption{Effect of attenuation factor ($\beta$) in HOPE on various tasks.}
	\label{fig:hope_param}
\end{figure*}
\begin{figure*}[!ht]
	\centering
	\subfloat[Graph Reconstruction]{\label{fig_m2} \includegraphics[width=0.32\textwidth]{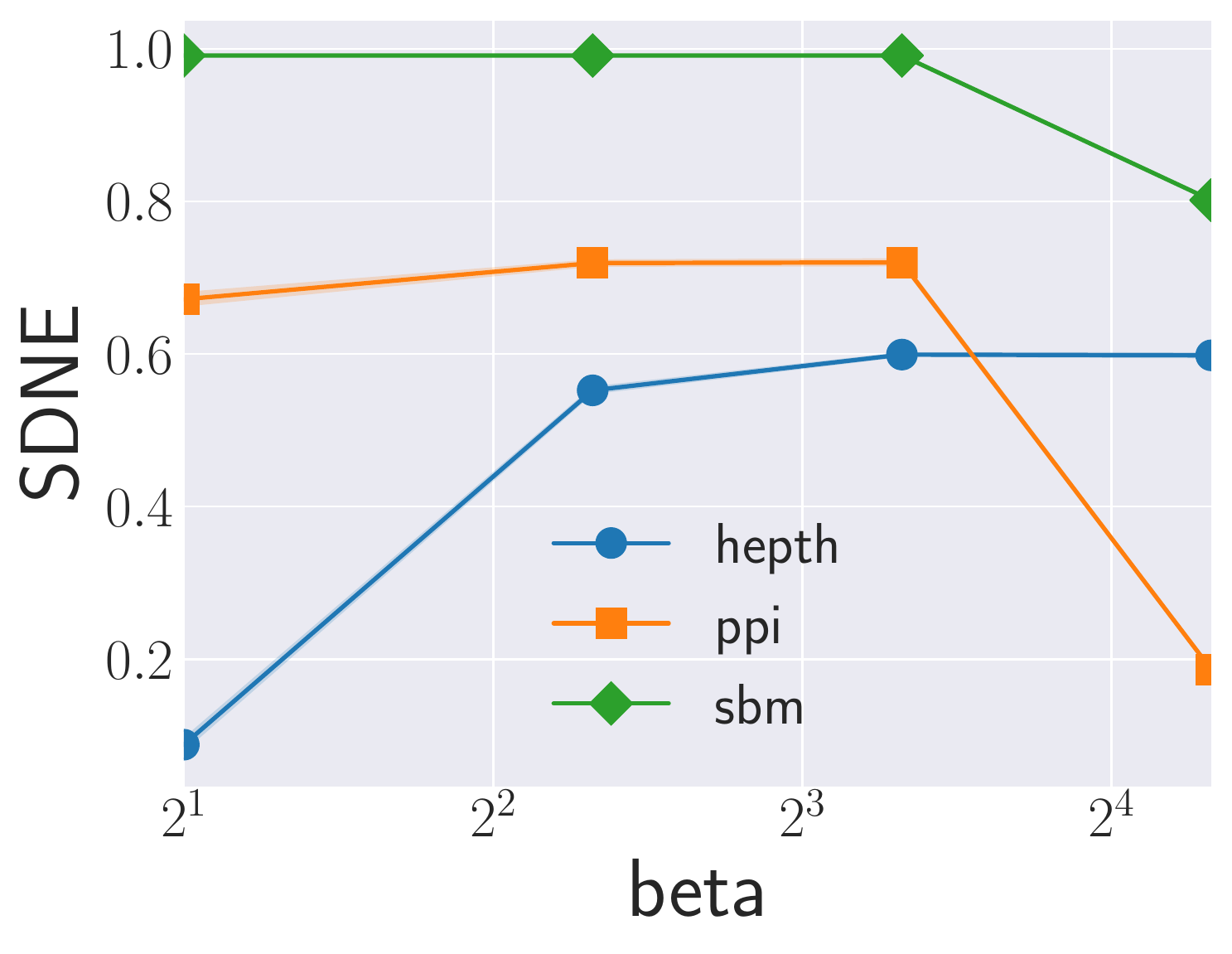}}
	\hfil
	\subfloat[Link Prediction]{\label{fig_m3} \includegraphics[width=0.32\textwidth]{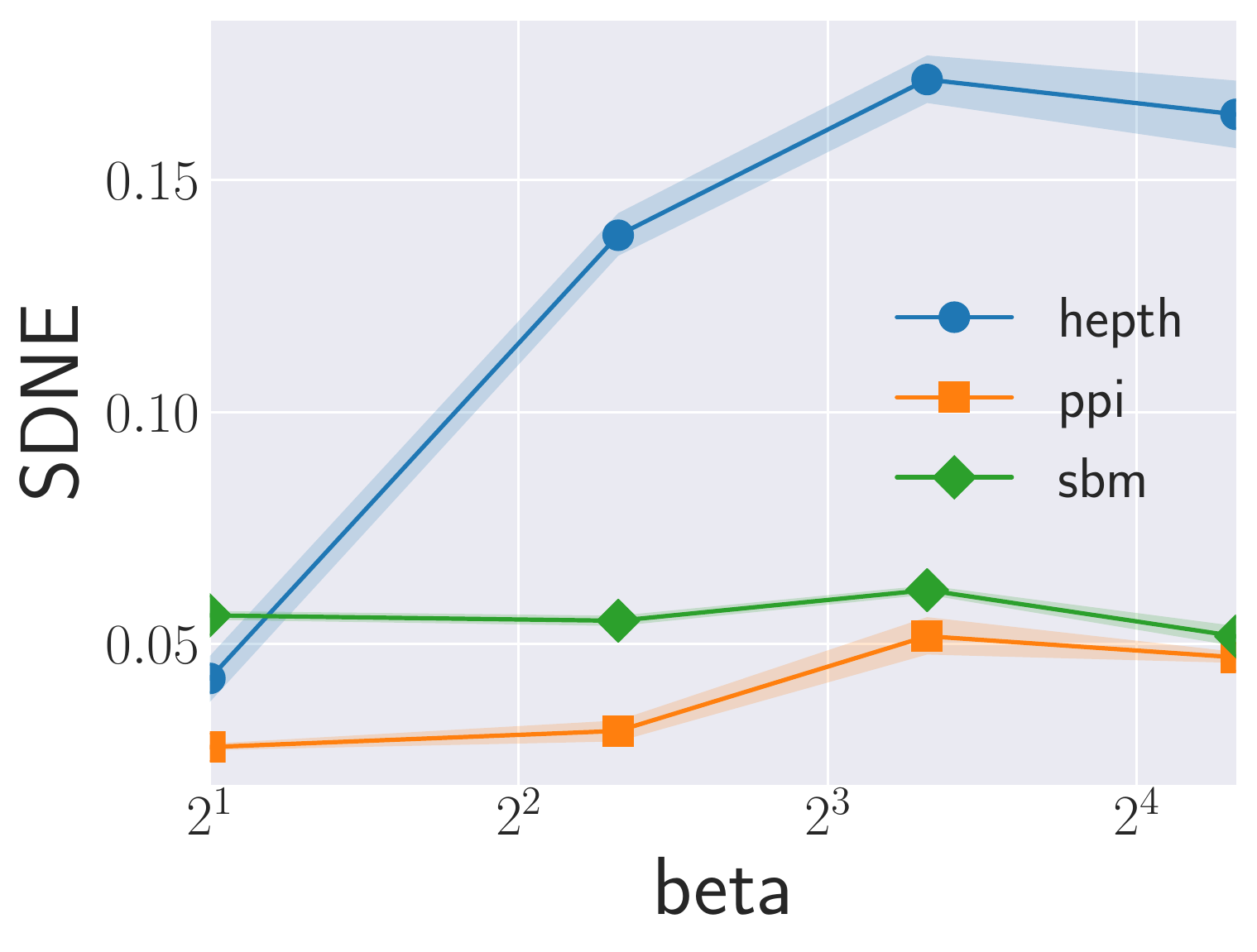}}
	\hfil
	\subfloat[Node Classification]{\label{fig_m4} \includegraphics[width=0.32\textwidth]{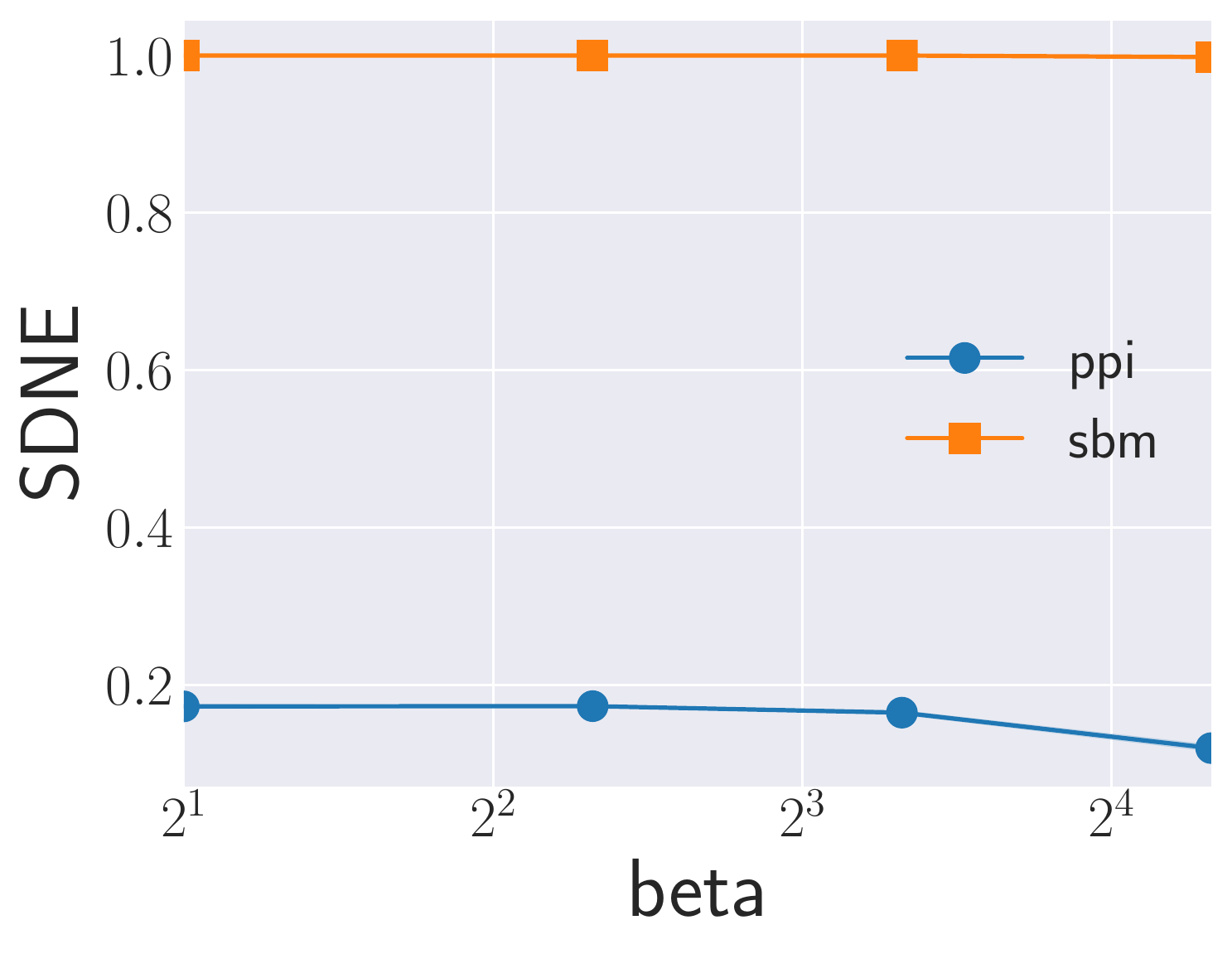}}
	\hfil
	\caption{Effect of observed link reconstruction weight in SDNE on various tasks.}
	\label{fig:sdne_param}
\end{figure*}
\begin{figure*}[!ht]
	\centering
    \subfloat[]{\label{fig_m2} \includegraphics[width=0.32\textwidth]{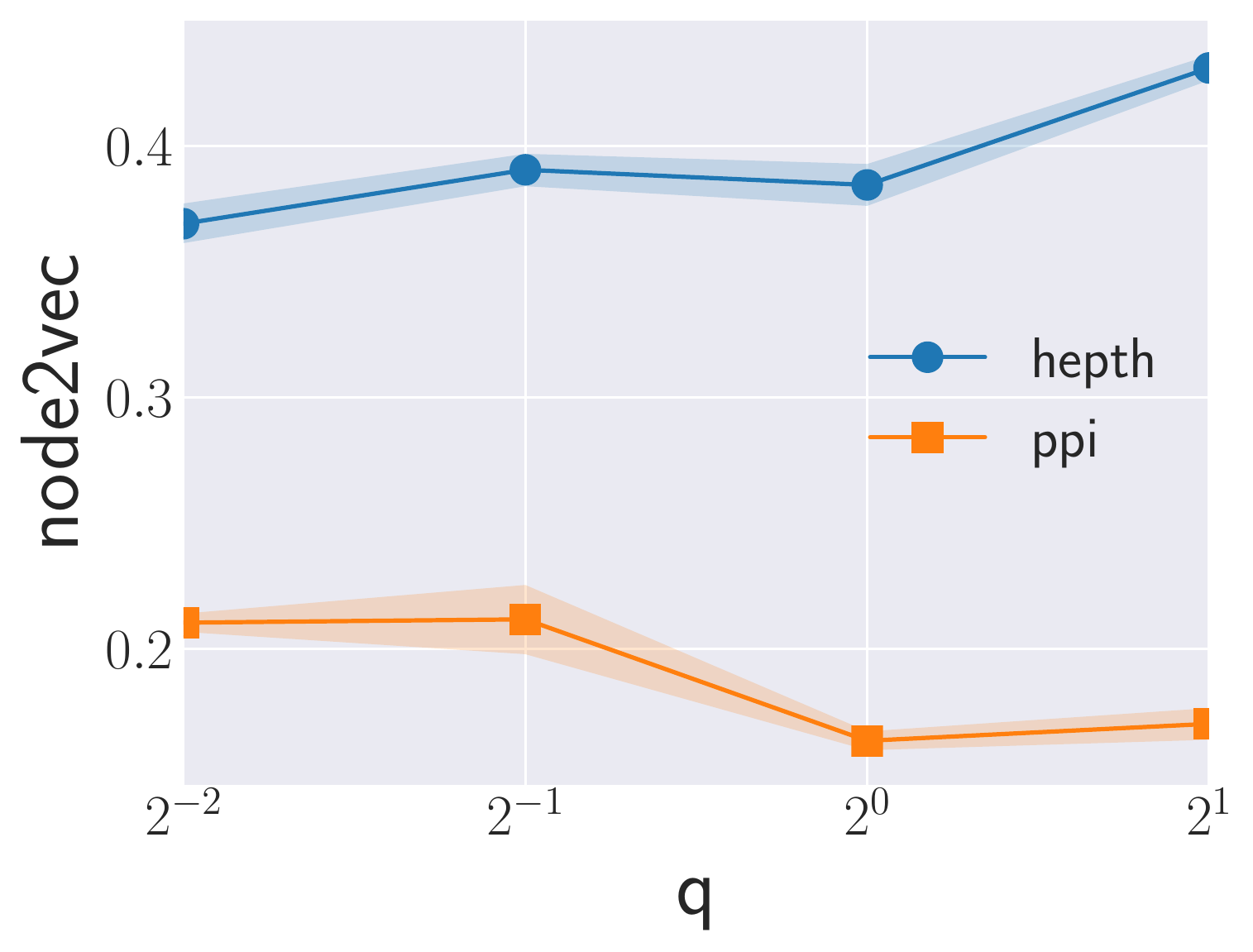}}
	\hfil
	\subfloat[]{\label{fig_m3} \includegraphics[width=0.32\textwidth]{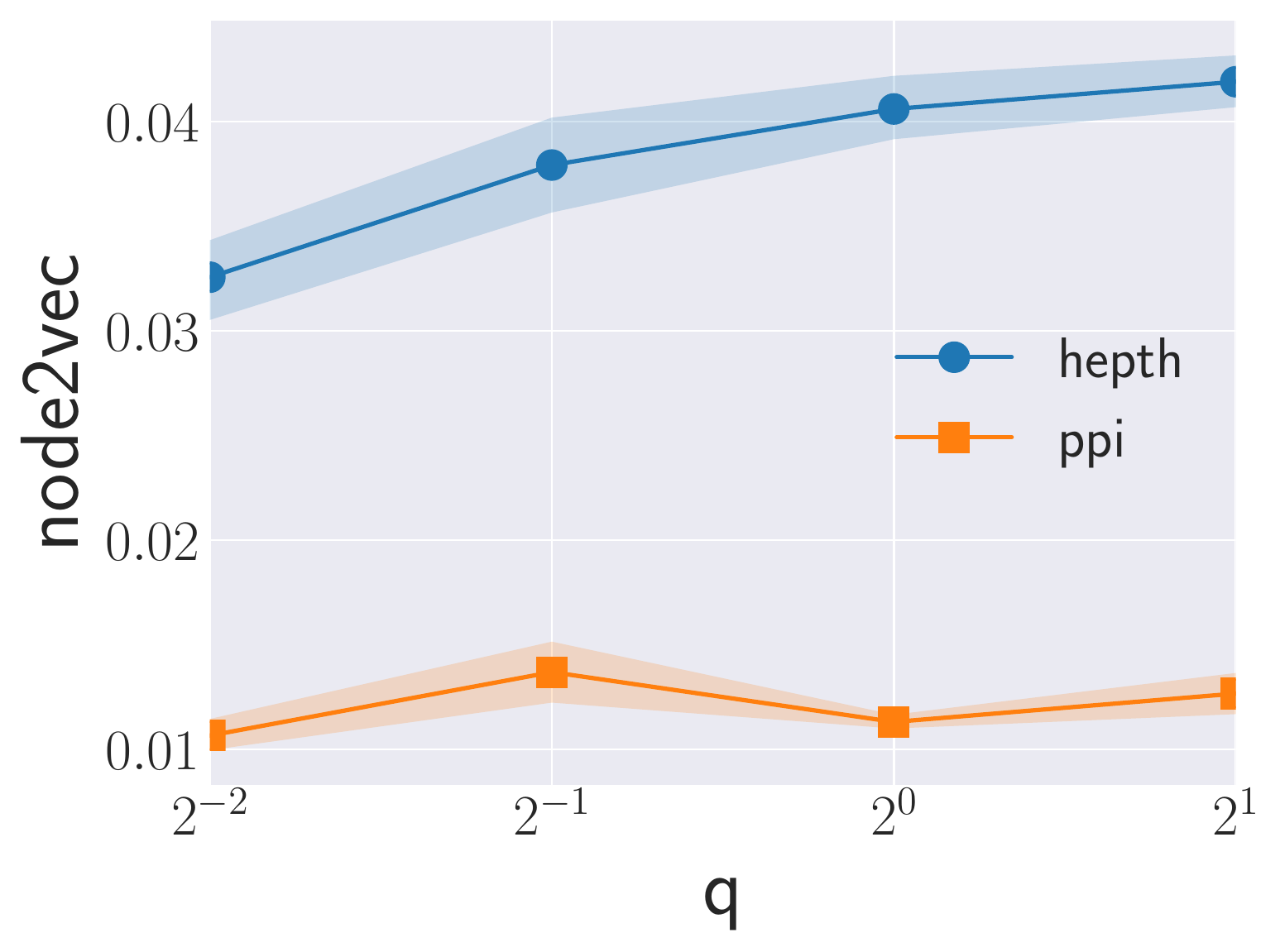}}
	\hfil
	\subfloat[]{\label{fig_m4} \includegraphics[width=0.32\textwidth]{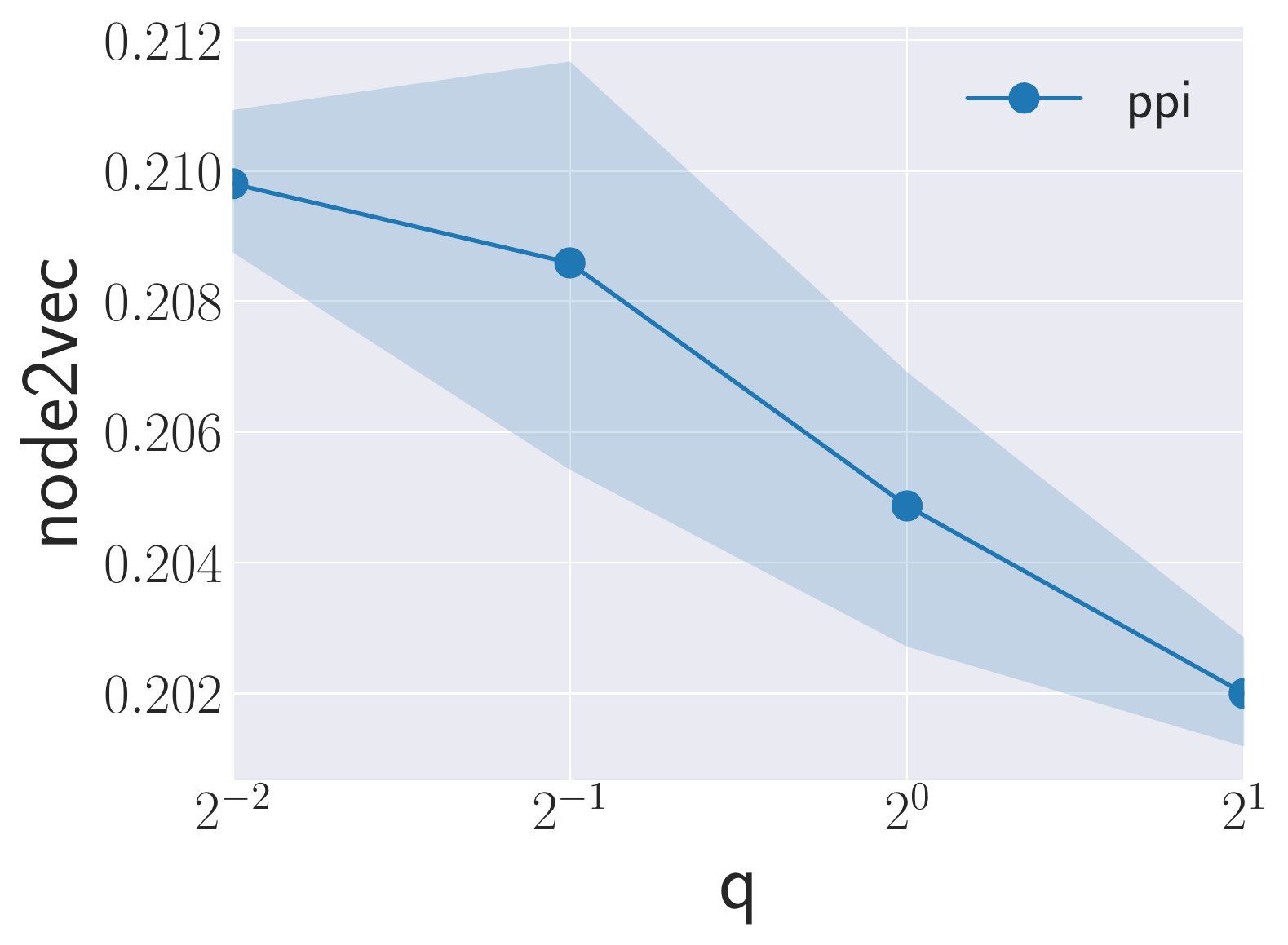}}
	\hfil
	\subfloat[Graph Reconstruction]{\label{fig_m2} \includegraphics[width=0.32\textwidth]{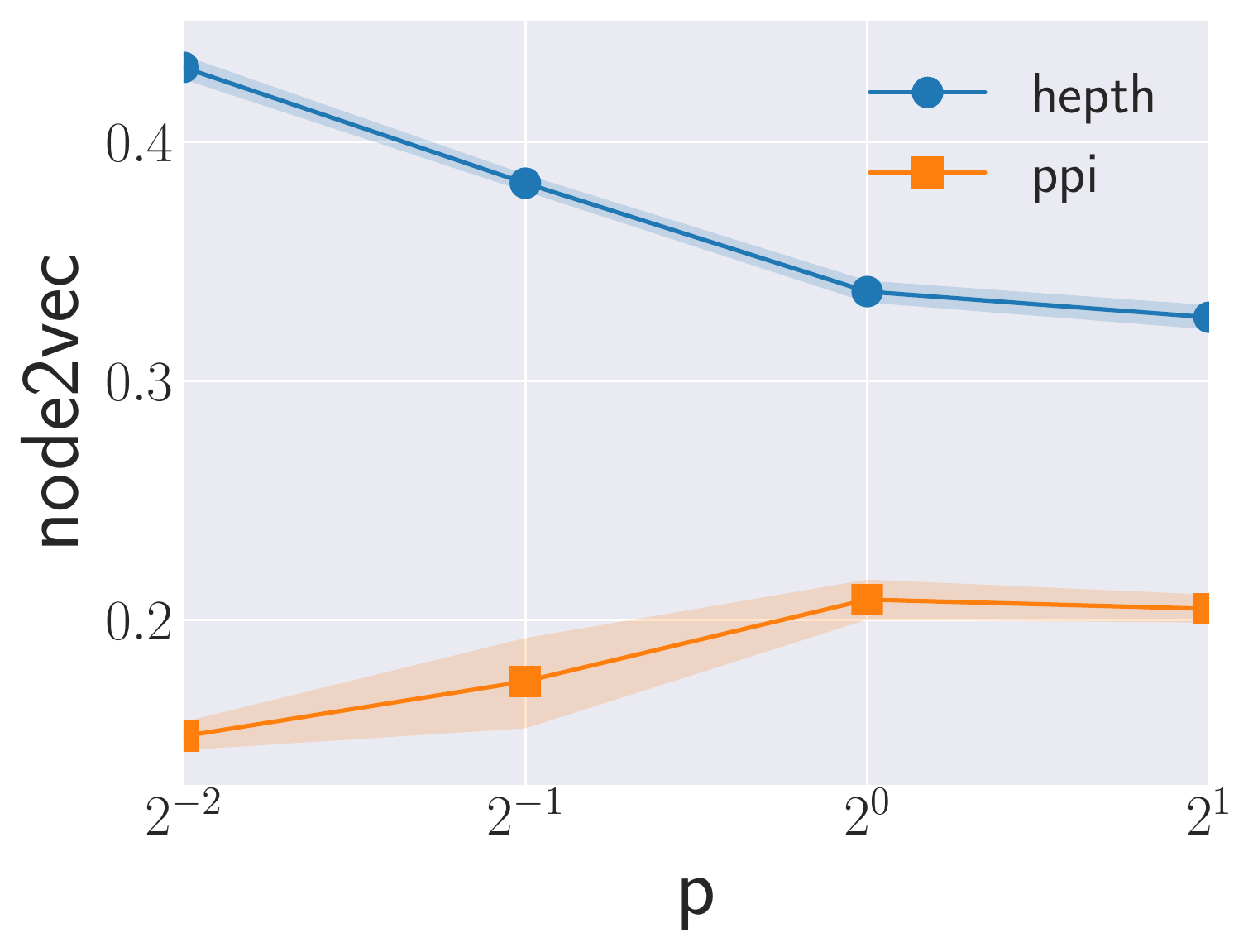}}
	\hfil
	\subfloat[Link Prediction]{\label{fig_m3} \includegraphics[width=0.32\textwidth]{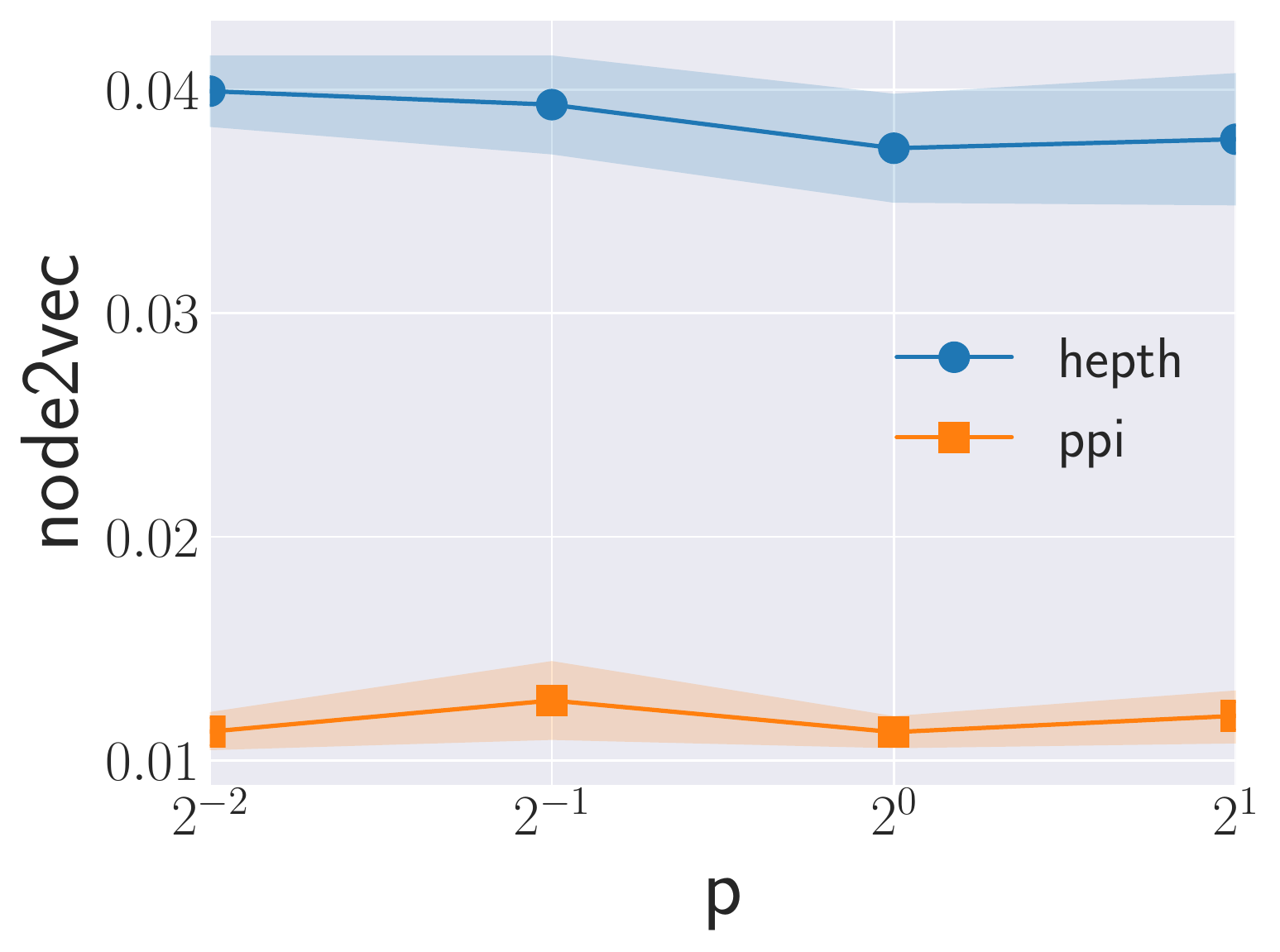}}
	\hfil
	\subfloat[Node Classification]{\label{fig_m4} \includegraphics[width=0.32\textwidth]{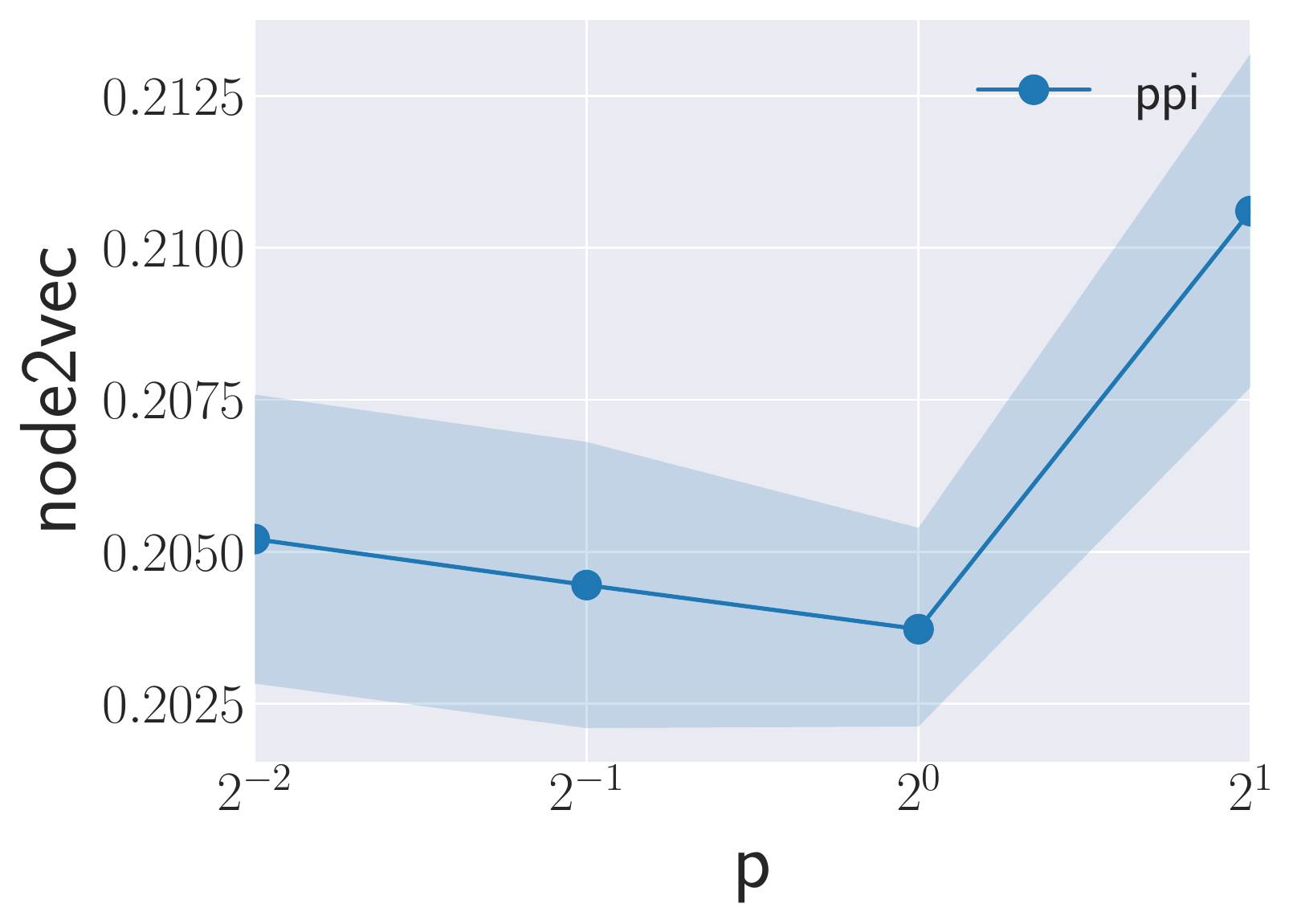}}
	\hfil
	\caption{Effect of random walk bias weights in \textit{node2vec} on various tasks.}
	\label{fig:node2vec_param}
\end{figure*}
\begin{figure}[!ht]
	\centering
    \subfloat[]{\label{fig_m2} \includegraphics[width=0.32\textwidth]{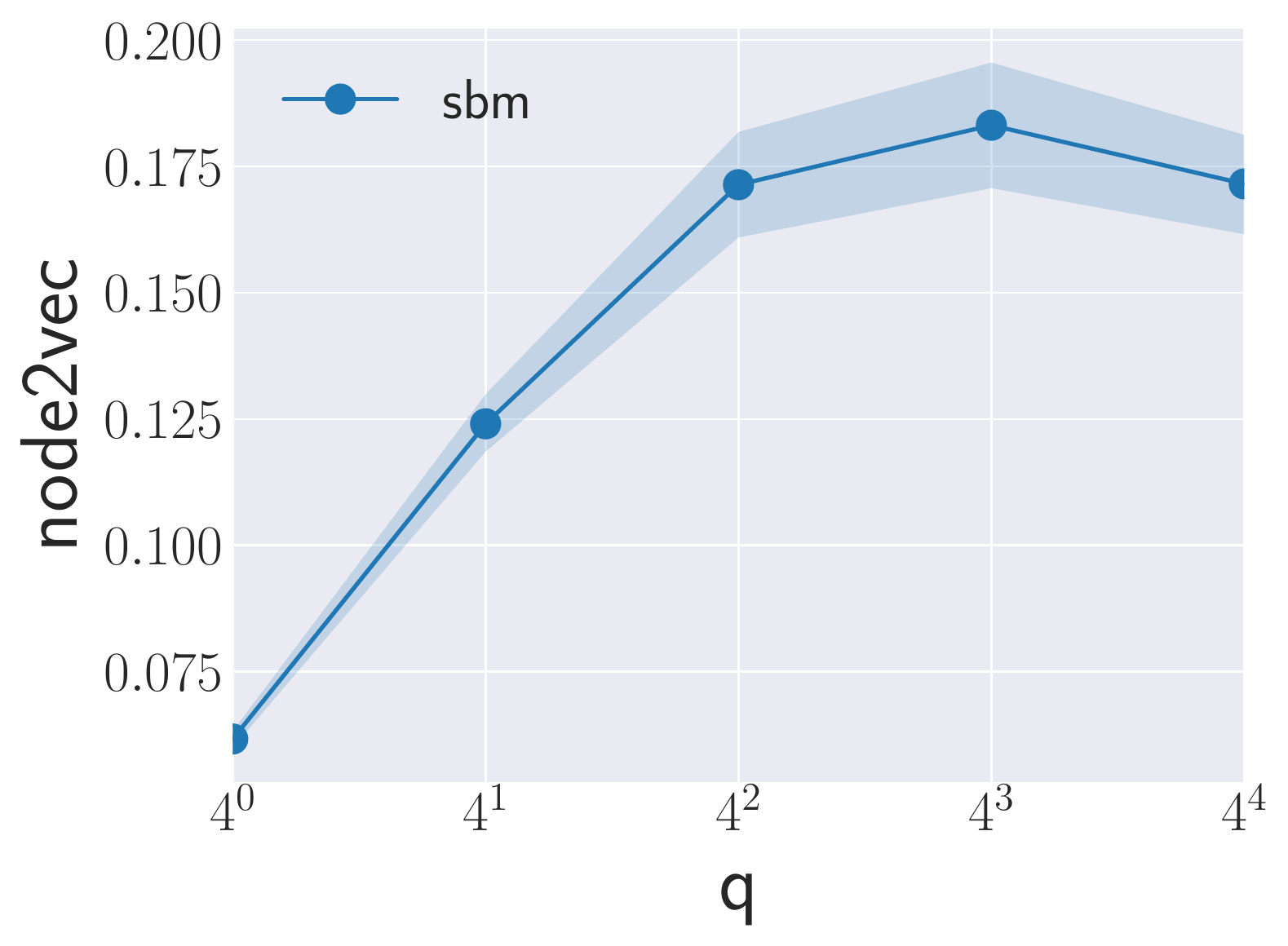}}
	\hfil
	\caption{Effect of random walk bias weights in \textit{node2vec} on SBM.}
	\label{fig:node2vec_param_sbm}
\end{figure}

In this section we plan to address the following questions:

\begin{itemize}
\item How robust are the embedding methods with respect to hyperparameters?
\item Do optimal hyperparameters depend on the downstream tasks the embeddings are used for?
\item What insights does performance variance with hyperparameter provide about a data set?
\end{itemize}

We answer these questions by analyzing performance of the embedding methods with various hyperparameters.
We present results on SBM, PPI and Hep-th.
Of the methods we evaluated in this survey, Laplacian Eigenmaps has no hyperparameters, thus is not included in the following analysis.

\textbf{Graph Factorization (GF)}. The objective function of GF contains a weight regularization term with a coefficient.
Intuitively, this coefficient controls the generalizability of the embedding.
A low regularization coefficient facilitates better reconstruction but may overfit to the observed graph leading to poor prediction performance.
On the other side, a high regularization may underrepresent the data and hence perform poorly on all tasks.
We observe this effect in Figure~\ref{fig:gf_param}.
We see that performance on prediction tasks, namely link prediction (Fig.~\ref{fig:gf_param}b) and node classification (Fig.~\ref{fig:gf_param}c), improves as we increase the regularization coefficient, reaches a peak and then starts deteriorating.
However, graph reconstruction performance (Fig.~\ref{fig:gf_param}a) may deteriorate with increasing regularization.
We also note that the performance change is considerable and thus the coefficient should be carefully tuned to the given data set.

\textbf{HOPE}. As HOPE factorizes a similarity matrix between nodes to obtain the embedding, the hyperparameters depend on the method used to obtain the similarity matrix.
Since in our experiments we used Katz index for this purpose, we evaluate the effect of the \textit{attenuation factor} ($\beta$), which can be interpreted as the higher order proximity coefficient on performance. 
Graph structure affects the optimal value of this parameter.
For well-structured graphs with tightly knit communities, high values of $beta$ would erroneously assign dissimilar nodes closer in the embedding space.
On the contrary, for graphs with weak community structure it is important to capture higher order distances and thus high values of $beta$ may yield better embeddings.
We validate our hypothesis in Figure~\ref{fig:hope_param}.
As our synthetic data SBM consists of tight communities, increasing $\beta$ does not improve the performance on any task.
However, gain in performance with increasing $beta$ can be observed in PPI and HEP-TH.
This is expected as collaboration and protein networks tend to evolve via higher order connections.
We observe that the optimal $beta$ for these data sets is lowest for the task of graph reconstruction (Fig.~\ref{fig:hope_param}(a)) and highest  for link prediction (Fig.~\ref{fig:hope_param}b), $2^{-6}$ and $2^{-4}$ respectively.
This also follows intuition as higher order information is not necessary for reconstructing observed edges but is useful for prediction tasks.
Nodes farther from each other are more likely to have common labels than links between them.
For example, in a collaboration network, authors in a field will form a weak community but many of them need not be connected to each other (in other words, most researchers of a same community won't be co-authors of one another).
Thus, if we were to predict field for an author, having information about other authors inside the weak community can help improve the accuracy of our model.

\textbf{SDNE}. SDNE uses a coupled deep autoencoder to embed graphs.
It leverages a parameter to control the weight of observed and unobserved link reconstruction in the objective function.
A high parameter value would focus on reconstructing observed links disregarding the absence of unobserved links.
This parameter can be crucial as a low value could hinder predicting hidden links.
Figure~\ref{fig:sdne_param} illustrates the results of our analysis.
We observe that performance on link prediction (Fig.~\ref{fig:sdne_param}b) greatly varies depending on the weight, with a more than 3-fold increase in MAP for Hep-th with the optimal parameter value, and about 2-fold increase for PPI.
The performance on node classification (Fig.~\ref{fig:sdne_param}c), on the other hand, is less affected by the parameter.
Overall, we see that the maximum performance is achieved for an intermediate value of the weight above which the performance drops.

\textbf{\textit{node2vec}}. \textit{node2vec} performs biased random walks on the graph and embeds nodes commonly appearing together in them close in the embedding space.
Of the various hyperparameters of the method (which include walk length, context size and bias weights), we analyze the effect of bias weights on performance and adopt commonly-used values for the remaining hyperparameter~\cite{Grover2016}, namely, context size of 10 and walk length of 80.
\textit{node2vec} has two bias weights: (a) \textit{inout} parameter ($q$), which controls the likelihood of random walk to go further away from the incoming node (higher values favor closer nodes), and (b) return parameter ($p$), which weighs the return probability (lower values favor return).
Lower values of $q$ would help capture longer distances between nodes and aims towards preserving structural equivalence.
Figure~\ref{fig:node2vec_param} illustrates the effect on PPI and HEPTH.
In node classification (Fig.~\ref{fig:node2vec_param}c) we observe that low $q$ values help achieve higher accuracy suggesting that capturing structural equivalence is required to accurately embed the structure of the graph.
On the contrary, high $q$ values favor link prediction (Fig.~\ref{fig:node2vec_param}b) following intuition that capturing community structure is crucial for the task.
We make similar observations for SBM in Figure~\ref{fig:node2vec_param_sbm} for the task of link prediction.
$MAP$ increases with increasing $q$ until it reaches an optimal.
We also note that the optimal values of $q$ in SBM are much higher as the graph has strong community structure.


\section{A Python Library for Graph Embedding}\label{sec:gem}
We released an open-source Python library, GEM (\textit{Graph Embedding Methods}, \url{https://github.com/palash1992/GEM}), which provides a unified interface to the implementations of all the methods presented here, and their evaluation metrics.
The library supports both weighted and unweighted graphs.
GEM's hierarchical design and modular implementation should help the users to test the implemented methods on new datasets as well as serve as a platform to develop new approaches with  ease.

GEM (\url{https://github.com/palash1992/GEM}) provides implementations of Locally Linear Embedding \cite{Roweis2000}, Laplacian Eigenmaps \cite{belkin2001laplacian}, Graph Factorization \cite{Ahmed2013}, HOPE \cite{Ou2016}, SDNE \cite{Wang2016} and node2vec \cite{Grover2016}.
For \textit{node2vec}, we use the C++ implementation provided by the authors \cite{snapnets} and yield a Python interface.
In addition, GEM provides an interface to evaluate the learned embedding on the four tasks presented above. 
The interface is flexible and supports multiple edge reconstruction metrics including cosine similarity, euclidean distance and decoder based (for autoencoder-based models). 
For multi-labeled node classification, the library uses one-vs-rest logistic regression classifiers and supports the use of other ad hoc classifiers.


\section{Conclusion and Future Work}\label{sec:conclusion}
This review of graph embedding techniques covered three broad categories of approaches: factorization based, random walk based and deep learning based.
We studied the structure and properties preserved by various embedding approaches and characterized the challenges faced by graph embedding techniques in general as well as each category of approaches.
We reported various applications of embedding and their respective evaluation metrics.
We empirically evaluated the surveyed methods on these applications using several publicly available real networks and compared their strengths and weaknesses.
Finally, we presented an open-source Python library, named GEM, we developed with implementation of the embedding methods surveyed and evaluation tasks including graph reconstruction, link prediction, node classification and visualization.

We believe there are three promising research directions in the field of graph embedding: (1) exploring non-linear models, (2) studying evolution of networks, and (3) generate synthetic networks with real-world characteristics.
As shown in the survey, general non-linear models (e.g. deep learning based) have shown great promise in capturing the inherent dynamics of the graph.
They have the ability to approximate an arbitrary function which best explains the network edges and this can result in highly compressed representations of the network.
One drawback of such approaches is the limited interpretability. 
Further research focusing on interpreting the embedding learned by these models can be very fruitful.
Utilizing embedding to study graph evolution is a new research area which needs further exploration.
Recent work by \cite{dai2017deep}, \cite{goyaldyngem} and \cite{zhu2016scalable} pursued this line of thought and illustrate how embeddings can be used for dynamic graphs.
Generating synthetic networks with real-world characteristics has been a popular field of research\cite{holland1983stochastic} primarily for ease of simulations.
Low dimensional vector representation of real graphs can help understand their structure and thus be useful to generate synthetic graphs with real world characteristics.
Learning embedding with a generative model can help us in this regard.

\section*{References}

\bibliography{mybibfile}

\end{document}